# An Energy Ontology for Global City Indicators (ISO37120)

**Alanna KOMISAR and Mark S. FOX**

**alanna.komisar@mail.utoronto.ca; msf@eil.utoronto.ca**

Enterprise Integration Laboratory (eil.utoronto.ca)
University of Toronto
5 King's College Road, Toronto ON, M5S 3G8

## 1. Introduction

To create tomorrow's smarter cities, today's initiatives will need to create measurable improvements. However, a city is a complex system and measuring its performance generates a breadth of issues including;

1. What criteria shall be measured and how would these indicators be defined?
2. How shall the identified indicators be derived?

In 2014, the ISO 37120 "Sustainable development of communities – Indicators of city services and quality of life" is an international standard published. With 100 defined indicators across 17 different themes, participating cites have consistent data to measure and compare their performance over time and against other municipalities.

To maintain consistency, the reported data is validated through a certification process organised by the World Council on City Data[1]. Although the certification process validates that the data has been collected in accordance with ISO37120, the process itself is closed from public view. A recent study shows that cities do not publish the data required for the public to verify the indicators (Fox & Pettit). Without the information on the indicators' derivation, the indicators will not provide enough information to understand the root causes of a city's performance.

This working paper is one in a series that addresses the creation of a Semantic Web-based representation of the 17 different themes of ISO37120 Indicators as part of the larger PolisGnosis Project (Fox, 2017; 2015). The objective of the PolisGnosis Project is to automate the longitudinal analysis (i.e. how and why a city's indicators change over time) and the transversal analysis (i.e. how and why the cities differ from each other at the same time), in order to discover the root causes of differences. A necessary condition for analysis is that both an indicator's definition and the data used to derive the indicator's value have to be represented using a standard representation and published on the semantic web. In this paper we define a standard representation/ontology for representing general knowledge for the Energy Theme indicators, and for representing both the definition and data used to derive the Energy indicators.

[1] http://www.dataforcities.org/global-cities-registry/

In the following we first specify a set of competency questions that the ontology must be able to answer. We then review the literature of energy indicators and ontologies. Followed by a review of the Global City Indicator Ontology that is part of the PolisGnosis project. We then define the GCI Energy Ontologies, followed the representation of the ISO37120 Energy Theme indicators definitions using these ontologies. We finish by evaluating the Energy ontologies using the competency questions.

# 2. Indicators and their competency requirements

## 2.1. ISO 37120 Energy Themed Indicators

Based on the methodology defined by Gruninger & Fox (1995), competency questions will be used to identify the types of information that the Energy ontology will need to represent in order define the Energy indicators and the data used to derive them. The following classifications will be used to organize the competency questions (Fox, 2015):

- **Factual (F):** Questions that ask what the value of some property is.
- **Consistency - Definitional (CD):** Questions that determine whether the instantiation of an indicator by a city is consistent with the ISO 37120 definition.
- **Consistency - Internal (CI):** Questions that determine whether different parts of the instantiation are consistent with each other.
- **Deduced (D):** A value or relationship that can be deduced from the instantiation.

The following meta-data related question apply to all indicators:

1. (F) What are the units of measure for the numerical value?
2. (F) When was the numerical value measured?
3. (F) Who or what agency measured the numerical value?
4. (F) What process was used to measure the value?
5. (CD) Is the indicator's supporting data consistent with the ISO37120 definition?

In the remainder of this section we provide a portion of the definition of each Energy theme indicator as found in the ISO 37120 standard. We then define a set of competency questions that the Energy ontology must support the answering of. These questions identify the various types of information the ontology must represent.

### 7.1 Total residential electrical energy use per capita (kWh/year) (Core Indicator)

**Indicator**: “Total residential electrical energy use per capita shall be calculated as the total residential electrical usage of a city in kilowatt hours (numerator) divided by the total population of the city (denominator). The result shall be expressed as the total residential electrical use per capita in kilowatt hours/year. Data should be gathered from electricity providers. Electricity consumption statistics are typically collected in three categories, residential, commercial and industrial.”

**Competency Questions**:

1.) (F) What city is the indicator for?
2.) (F) What is the total population of the city?
3.) (F) Is "Toronto Building 01" a residential building?
4.) (CD) Who are the owners of "Toronto Building 01"? What sector owns this building?
5.) (F) What percentage of the floor space is used for residential purposes?
6.) (F) How much energy was used per year in residential buildings?
7.) (F) What organizations provide electrical service in Toronto?
8.) (F) What addresses does "Service Provider A" service?
9.) (F) How many service accounts are there in "Toronto Building 01"?

### 7.2 Percentage of city population with authorized electrical service (Core Indicator)

**Indicator:** "The percentage of city population with authorized electrical service shall be calculated as the number of persons in the city with lawful connection to the electrical supply system (numerator) divided by the total population of the city (denominator). The result shall then be multiplied by 100 and expressed as a percentage.

The number of city households lawfully connected to the electricity grid shall be multiplied by the current average city household size to determine the number of city residents with lawful connection to the electricity supply system (the electricity grid)"

**Competency Questions**:

1.) (F) How many electrical service accounts are there in the city?
2.) (F) How many households does service account x provide energy to?
3.) (F)How many households are in the city?
4.) (F) How many households in total are legally serviced by these electrical accounts?
5.) (F) What is the average number of people living in each household?
6.) (F) How many households does "Toronto Building 01" have?
7.) (D) How many individual electrical service account holders hold more than one account?

### 7.3 Energy consumption of public buildings per year (kWh/m²) (Core Indicator)

**Indicator:** "Energy consumption of public buildings shall be calculated per year as the total use of electricity at final consumption stage by public buildings (kWh) within a city (numerator) divided by total floor space of these buildings in square meters (m²) (denominator). The result shall be expressed as the total energy consumption of public buildings per year in kilowatt hours per square meter. Note that public buildings are government owned buildings such as government offices, hospitals and schools."

**Competency Questions:**

1.) (F) Which buildings in Toronto are owned by the government?
2.) (F) How many public buildings are in the city?

3.) (F) What is the floor area of all the public buildings?
4.) (F) How many kWh are being consumed by public buildings?

### 7.4 The percentage of total energy derived from renewable sources, as a share of the city's total energy consumption (Core Indicator)

**Indicator**: "The share of a city's total energy consumption derived from renewable sources shall be calculated as the total consumption of electricity generated from renewable sources (numerator) divided by total energy consumption (denominator). The result shall then be multiplied by 100 and expressed as a percentage. Consumption of renewable sources should include geothermal, solar, wind, hydro, tide and wave energy, and combustibles, such as biomass.

Data available from local utility provider, city energy or environment office, and from various international sources, such as the International Energy Agency (IEA), and the World Bank.

Renewable energy shall include both combustible and non-combustible renewables. Non-combustible renewables include geothermal, solar, wind, hydro, tide and wave energy. For geothermal energy, the energy quantity is the enthalpy of the geothermal heat entering the process. For solar, wind, hydro, tide and wave energy, the quantities entering electricity generation are equal to the electrical energy generated. The combustible renewables and waste (CRW) consist of biomass (fuel wood, vegetal waste, ethanol) and animal products (animal materials/waste and sulphite lyes), municipal waste (waste produced by the residential, commercial and public service sectors that are collected by local authorities for disposal in a central location for the production of heat and/or power) and industrial waste."

**Competency Questions**:

1.) (F) What is the total electrical usage of the city?
2.) (F) Who are the electrical service providers for the city?
3.) (F) What is the electrical mixture distributed for each provider?
4.) (F) What are the different sources of electricity generation?
5.) (F) Which sources are renewable? Non-renewable?
6.) (F) Which sources are combustible renewable?
7.) (F) What is the total enthalpy for all geothermal heat?
8.) (F) What % of the usage is fossil fuels?
9.) (F) Who provided the information on the renewable energy mixture production?

### 7.5 Total electrical energy use per capita (kWh/year) (supporting indicator)

**Indicator**: "Total electrical energy use per capita shall be calculated as the total electrical usage of a city in kilowatt hours including residential and non-residential use (numerator) divided by the total population of the city (denominator). The result shall be expressed as the total electrical use per capita in kilowatt hours/year.

Data shall be gathered from electricity providers. Electricity consumption statistics are typically collected in three categories: residential, commercial and industrial. Compilation of the sources used to generate energy based on fossil and renewable energy sources; types of renewable energy already in use; identification of locally existing renewable energy sources; compilation of the energy required for heating and cooling processes; completed and planned measures to save energy and to improve energy efficiency; completed and planned activities for the environmentally friendly insulation and cooling of buildings, if available should be noted."

**Competency Questions:**

1.) (F) What is the total electrical usage per year?
2.) (F) Who reported the total electrical usage?
3.) (CI) What is the total non-residential energy usage per year?
4.) (CI) What is the total residential energy usage per year?

### 7.6 Average number of electrical interruptions per customer per year (supporting indicator)

**Indicator**: "The average number of electrical interruptions per customer per year shall be calculated as the total number of customer interruptions (numerator) divided by the total number of customers served (denominator). The result shall be expressed as the average number of electrical interruptions per customer per year.

Electrical interruptions shall include both residential and non-residential. It is normal to expect interruptions in service for a number of reasons including scheduled maintenance and equipment breakdown. To establish the opportunity to have a reasonable comparison between energy providers, major storms and weather events should be excluded due to their unpredictability and randomness since they are difficult to predict, prevent or mitigate against. This indicator is affected by the age, standard of maintenance and reliability of the infrastructure that constitutes the electricity grid and the electricity transmission capacity that services the grid. The ability of both the grid and its electricity transmission capacity to provide supply on demand and to cope with peak loads is also an important consideration."

**Competency Questions:**

1.) (CD) How do you define customers? Is that number of households or as the number of individuals under assumed impact?
2.) (CD) If an individual's home and office gets an electrical interruption, will they be counted twice?
3.) (F) How many Households have electrical interruptions?
4.) (F) How many non-residential electricity users were impacted by the electrical interruption?
5.) (F) How many service interruptions were caused by extreme weather events?
6.) (F) How many interruptions were there in the city total?
7.) (F) Which account holders total were impacted by the interruption?
8.) (F) Which electrical providers experiences the most interruptions?

9.) (F) How many interruptions were counted due to extreme weather events?

### 7.7 Average Length of electrical interruption (hours) (supporting indicator)

**Indicator:** “The average length of electrical interruptions shall be calculated as the sum of the duration of all customer interruptions in hours (numerator) divided by the total number of customer interruptions (denominator). The result shall be expressed as the average length of electrical interruptions in hours. Electrical interruptions shall include both residential and non-residential. It is normal to expect interruptions in service for a number of reasons including scheduled maintenance and equipment breakdown. To establish the opportunity to have a reasonable comparison between energy providers, major storms and weather events shall be excluded due to their unpredictability as they are difficult to prevent or mitigate against.

This indicator is affected by the age, standard of maintenance and reliability of the infrastructure that constitutes the electricity grid and the electricity transmission capacity that services the grid. The ability of both the grid and its electricity transmission capacity to provide supply on demand and to cope with peak loads is also an important consideration.”

**Competency Questions**:

1.) (F) What is the total duration of all electrical interruptions?
2.) (F) What is the total number of interruptions?

# 3. Background

This section reviews the ontologies that were investigated and re-used in the development in the GCI Energy ontology. The ontologies evaluated broadly fit into two classes; ontologies that pertain to energy usage and ontologies that have terminology that could be reused to structure where and how energy is used in a city.

## 3.1. Energy Ontologies and Resources

Although there are many ontologies and resources on the topic of “Energy” available, the existing choices did not satisfy the competency questions specified for the GCI Energy Themed Ontology. Most existing energy ontologies are developed to facilitate improving energy efficiency on a specific building level or for smart grid infrastructure rather than total energy usage at a municipal level. These ontologies will be discussed later in this section

Concepts for the sources of energy generation were drawn from the Semanco Urban Energy Ontology (sc) (Sicillia et al., 2014). Semanco is based on existing knowledge standards on building energy efficiency and it has concepts for economic, climate, and social factors that impact energy usage. Semanco was constructed to describe the following knowledge standards:

- ISO/ICE CD 13273 –Energy Efficiency and Renewable Energy Sources
- ISO/DTR 16344 – Common Terms, Definitions and Symbols for the overall Energy Performance Rating and Certification of Buildings
- ISO/CD 16346 – Assessment of Overall Energy Performance of Buildings
- ISO/DIS 12655 – Presentation of real energy use of buildings

- ISO/CD 16343 – Methods for expressing energy performance and for energy certification of buildings
- ISO 5001:2011 – Energy Management Systems – Requirements with Guidance for Use

However, ISO 37120 was not one of the knowledge standards included in this list. Semanco represents buildings on a single building level rather than at a municipality level. It also does not have information on electrical interruptions. From this ontology, terminology on both renewable and non-renewable energy sources is reused in the GCI energy ontology, such terms include “Not-Renewable_Energy_Source,” “Nuclear,” “Fossil_Fuels,” “Renewable_Energy_Source,” “Solar_Energy,” “Wind_Energy,” “Geothermal_Energy,” “Biomass,” and “Hydro_Energy.”

Although the Semanco was the primary ontology where energy related terms were derived from, numerous other energy related ontologies that investigated but not used for the construction of the GCI Energy Ontology for two reasons; many required concepts existed in Semanco; and second, the GCI Energy ontology explores macro level trends and concepts of a city whereas many of these ontologies explored micro-level data such as the energy output of individual home automation devices or solar panels. These ontologies include:

- OntoEnergy - Energy Efficiency for Manufacturing Plant Automation - (Linnenberg, Christiansen, & Seitz, 2013)
- DNAS - Behavioural Psychology of Energy Usage in Buildings – (Hong et al., 2015)
- A Generic Ontology for Prosumer Oriented Smart Grids- (Gillani, Laforest, & Picard, 2014)
- Smart Grid Energy Management Systems (Zeiler & Boxem, 2013)
- Critical Infrastructure Ontology – (Masucci, Adiolfi, Servillo, & Tofani, 2009)
- OpenWatt Renewable Energy Data Sources (Lamanna, Davide, Maccioni, 2012)
- DEHEMs - Digital Environment Home Energy Management System DEHEMs (Tommis, 2011)

## 3.2 Non-Energy Related Ontologies and Resources

### 3.2.1 Service

Additional non-energy related ontologies were used to build the GCI Energy Ontology to structure the classifications of buildings, how a building interacts with its electrical service, and general information about city populations.

The framework of the GCI Energy Sub-Ontologies was based off of the structure created in the GCI Telecommunications and Innovation Ontology (Forde & Fox, 2015). The ontology contains two sub-ontologies; “Service” and “Residency.” These ontologies provided the basic framework for the energy ontology because electrical services have similarities to telecommunications on how they are purchased and distributed. Although the GCI Telecommunications and Innovation Ontology provides many necessary terms, to answer the competency questions for GCI Energy, the “Service” and “Residency” ontologies do not include electricity related terms, building classes (beyond residential), nor track service interruptions.

Most of the service related concepts used in the Telecommunications and Innovation Ontology came from Service Micro-Ontology (Voß, 2013) such as “Service,” “ServiceProvider,” and “ServiceConsumer,” “ProvidedBy” and “ConsumedBy.”

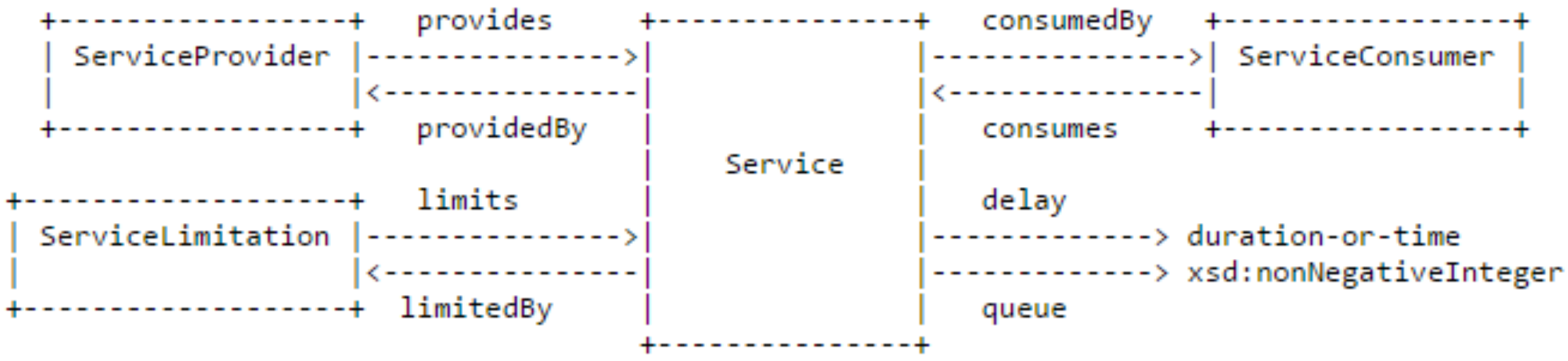

*Figure 1 SO Service Ontology*

Beyond the terms found in the Service Ontology by Voß, the term “APurchase” was imported directly from the GCI Innovation and Telecommunication ontology to define a legal transaction between a service consumer and service provider that would authorize a legal connection to the services.

### 3.2.2 Populations and Households

The classes that define people and users and electrical services were drawn from the Organization Ontology (Fox, 1998), The GCI Shelter Ontology (Wang & Fox, 2015), the FOAF Ontology (Miller, 2014).

From the Organization Ontology, the terms “Organization,” “Division,” “Government Organization,” and “Non-Government Organization” were imported to describe the service user groups living in the city and owning the services and buildings.

The residential population in the ISO 37120 indicators is described as a function of the average household size for the city. The terms for “household,” “household_size,” “average_household,” and “average_household_size” were imported from the GCI Shelter Ontology.

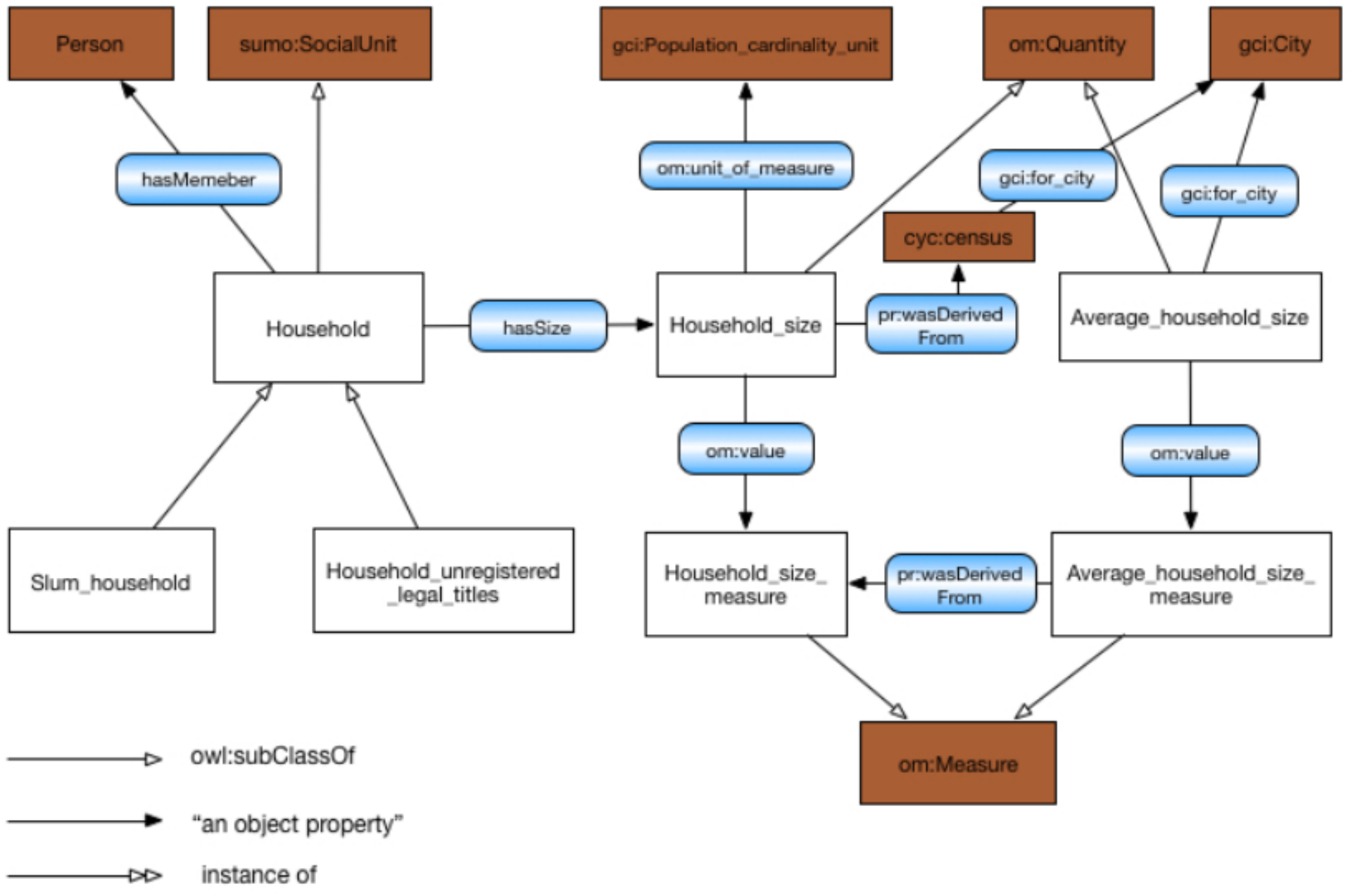

*Figure 2 GCI Shelter Ontology Household Definition*

## 3.3 GCI Foundational Ontology

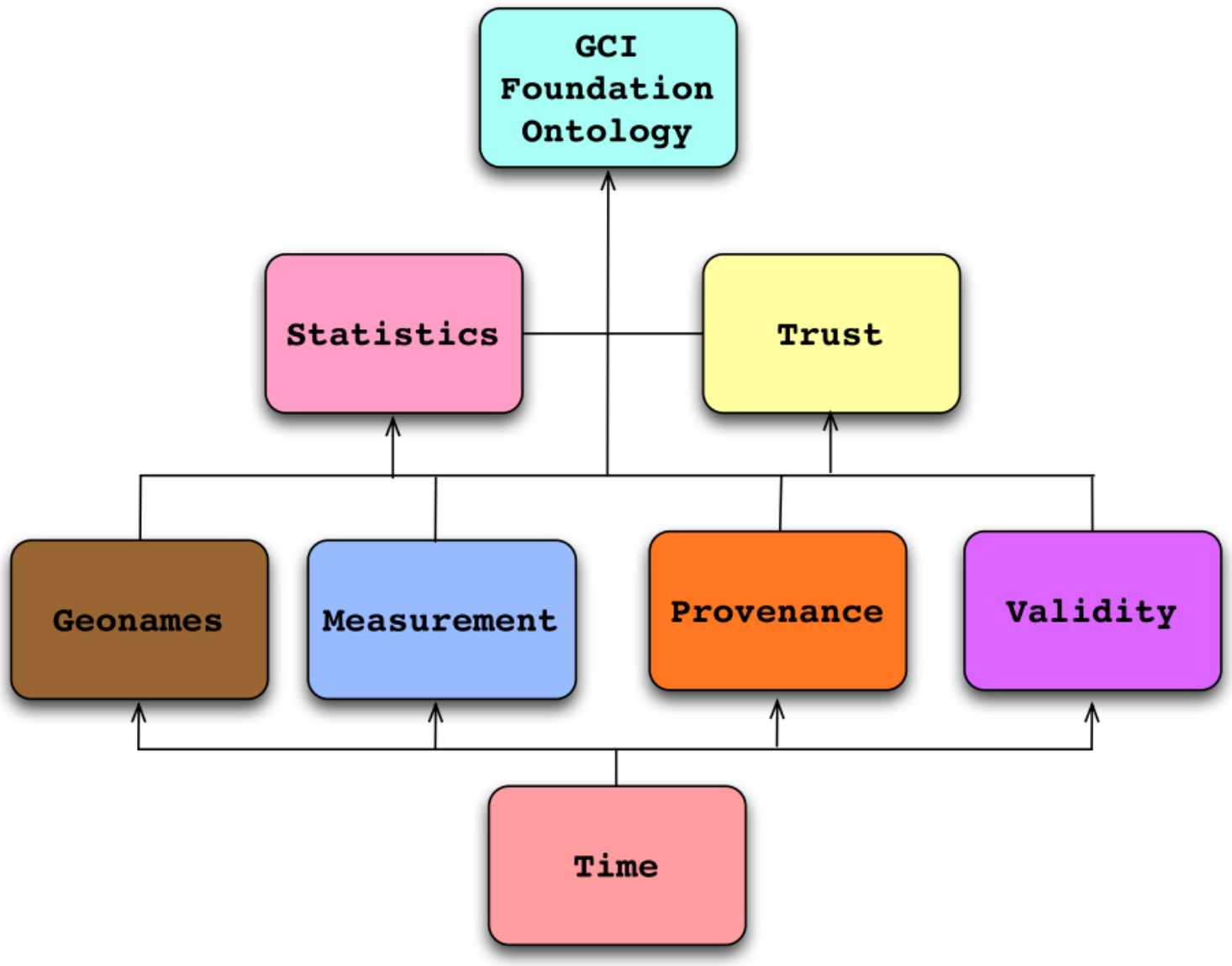

*Figure 3 GCI Foundational Ontology*

The GCI Foundation Ontology (Fox, 2013) provides concepts and properties that are necessary to represent all ISO 37120 indicators. It defines the representation of meta-information associated with a single indicator number, including: place names, units, time, provenance, validity and trust.

It also defines the classes and properties for representing the definition of an indicator, including populations, how they are measured and how they are analytically combined within an indicator. The foundation ontology integrates and extends the following ontologies:

- Time (Pan & Hobbs, 2004)
- Measurement (H. Rijgersberg, 2011)
- Statistics (Pattuelli, 2009)
- Geonames Ontology (www.geoname.org)
- Provenance (Belhajjame et al., 2012)
- Validity (Fox & Huang, 2005)
- Trust (Huang & Fox, 2006)

## 4. Architecture of the Global City Indicator Ontology

As explained in the introduction of this paper, ISO 37120 defines 100 city indicators. The following diagram (Fox, 2013) depicts the modules that are used to represent the definitions of the 37120 indicators. The ISO37120 module provides the IRIs for uniquely identifying all 100 indicators. For example, the IRI for the Total Electricity Usage Per Capita indicator is: http://ontology.eil.utoronto.ca/ISO37120.owl#7.5.

Each ISO37120 theme's indicators are defined in separate files. The OWL representation of the definitions of all seven of the GCI Energy indicators can be found in Energy.owl.

The GCI Ontology level provides specific ontologies required to define each theme's indicators. The Energy indicators are defined with concepts such as service, buildings, service interruptions, etc.  These concepts are captured in GCIBuildingOccupancy.owl and GCIService.owl and are used in Energy.owl.

All of the ontologies specifics to individual themes are built on base on GCI Foundation ontology, which defines generic concepts such as energy units, meta-information, etc.

The Enterprise Ontology level contains the TOVE Enterprise Modelling ontologies (Grüninger, 1998). In this figure, only the Organization Ontology File[2] (Fox et al., 1998), one of the TOVE Enterprise Modelling ontologies, is shown.

Lastly, the Foundation Ontology level contains the very basic ontologies which serve as the foundation for everything above.

[2] The Organization ontology can be found at http://ontology.eil.utoronto.ca/organization.owl along with its documentation at http://ontology.eil.utoronto.ca/organization.html. The prefix "org" is used for this ontology where needed.

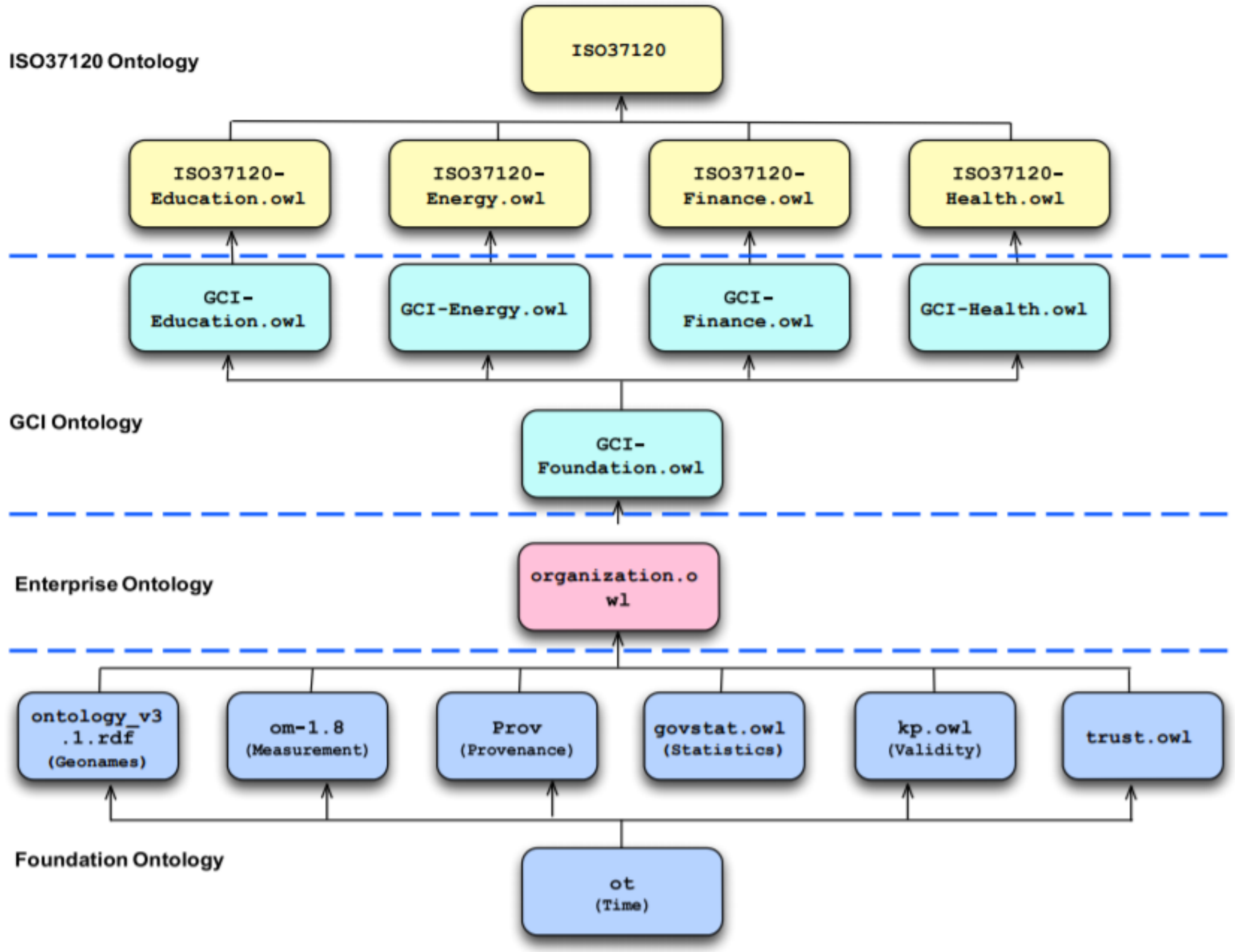

*Figure 4 GCI Foundational Ontology Architecture*

# 5. GCI Energy Sub-Ontologies

## 5.1 GCI - Service

Electricity is distributed as a service in cities, so in order to construct an ontology for ISO37120 energy indicators, a service ontology is needed. To satisfy the ISO37120 energy indicators competency questions, information on the quantity of the service used, the number of people with access to the service, frequency and duration on interruptions, and source of the services will need to be constructed. Some of the competency questions that relate to service include:

1. (F) How much electricity was used per year in residential buildings?
2. (F) What organizations provide electrical service in Toronto?
3. What addresses does "Service Provider A" service?
4. (F) How many service accounts are there in "residential building 01"?
5. (F) How many electrical service accounts are there in the city?
6. (F) How many households does service account x provide energy to?
7. (D) How many individual electrical service account holders hold more than one account?
8. (F) How many kWh are being consumed by public buildings?
9. (F) What is the total electrical usage of the city?

10. (F) What percentage of the electricity distributed comes from renewables by service provider?
11. (F) Who provided the information on the renewable energy mixture production?
12. (F) How many Households have electrical interruptions?
13. (F) How many interruptions were counted due to extreme weather events?
14. (F) What is the total duration of all electrical interruptions?
15. (F) What is the total number of interruptions?

This service ontology extends the Document Service Ontology (prefix so) (Voß, 2013) and the GCI's Innovation Ontology (prefix gcii) (Forde & Fox, 2015) to include information about the total volume of service used, views at the account and building level, service interruptions, and accounts that service multiple tenants.

| Class | Property | Value Restriction |
|---|---|---|
| Service | owl:subclassOf | so:Service |
| | distributedBy | min 1 so:ServiceProvider |
| Service_Measure | rdf:subclassOf | gci:GCI_measure |

**Authorized Service Connections**

A service consumer is an individual who has an authorized service connection. In the case of many different types of services, a single service account will benefit a group who lives or works at the address that it is servicing. This is described by ServiceConsumerGroup, where this can be defined as a household, organization, or a division of a greater organization.

| Class | Property | Value Restriction |
|---|---|---|
| ServiceConsumer | owl:subclassOf | so:ServiceConsumer |
| | owl:subclassOf | foaf:agent |
| | experienceService Interruptions | only ServiceInterruption |
| | so:consumes | min 1 Service |
| | connectedthrough | min 1 ServiceAccount |
| | authorizedBy | min 1 ServiceProvider |
| ServiceConsumerGroup | consist_of | min 1 ServiceConsumer |
| | experienceService Interruption | only ServiceInterruption |
| | authorizedBy | min 1 ServiceProvider |
| | so:consumes | min 1 Service |
| | using | min 1 serviceAccount |
| | legallyAuthorizedBy | min 1 gcii:APurchase |
| | represents_a | some (org:Division or gcis:Household or org:Organization) |
| ServiceProvider | owl:subclassOf | so:ServiceProvider |
| | owl:subclassOf | org:Organization |
| | authorizes | min 1 ServiceAccount |
| | so:provides | min 1 Service |
| | ic:hasAddress | some ic:Address |

The class ‘APurchase’ is from gcii. It includes properties from schema.org and so. From schema.org the class ‘Offers’ is used in which ‘APurchase’ is a subclass. An ‘Offer’ is defined as, “the transfer of some rights to an item or to provide a service.” ‘APurchase’ is a member of the Offer class since it forms a transaction between the ‘ServiceProvider’ and ‘ServiceConsumer’. As in the GCI Innovation Ontology, if a consumer has access to “APurchase” it defines that a legally authorized connection has been made.

| Class | Property | Value Restriction |
|---|---|---|
| gcii:APurchase | owl:subclassOf | sch:Offer |
| | gcii:consumedBy | some so:ServiceConsumer |
| | gcii:providedBy | exactly 1 so:ServiceProvider |
| | gcii:servicetype | exactly 1 so:Service |
| | gcii: ‘price currency’ | some xsd:decimal |
| | gcii:certification_Date | exactly 1 xsd:dateTime |
| | gcii:expiry_Date | exactly 1 xsd:dateTime |

The term ServiceAccount is introduced as a subclass of APurchase. It authorizes a legal connection between a service provider and a consumer. The definition was extended to indicate whether the account is currently active and also provide measurement of the total amount of the service that is consumed. It also addresses that a single transaction can be connected to multiple addresses.

| Class | Property | Value Restriction |
|---|---|---|
| ServiceAccount | owl:subclassOf | gcii:APurchase |
| | accountActive | exactly 1 xsd:boolean |
| | authorizedBy | exactly 1 ServiceProvider |
| | providedBy | exactly 1 ServiceProvider |
| | hasServiceAddress | min 1 ic:Address |
| | hasServiceArea | only gcis:ServiceAreaMeasure |
| | hasServiceType | only Service |
| | consumedBy | min 1 ServiceConsumer |
| | owned_by | min 1 foaf:Agent |
| | hasConsumption | min 1 gci:GCI_quantity |
| | hasServiceInterruption | only ServiceInterruption |

**Service Consumption**

To measure the total amount of service consumed by sectors, accounts, consumers, and individual buildings, Service_Measure and Service_Consumption_Quantity were created. Service_Consumption_Quantity defines the numerical value of service consumed by either a service consumer or a service consumer group.

| Class | Property | Value Restriction |
|---|---|---|
| Service_Consumption_Quantity | owl:sublclassof | gci:GCI_quantity |
| | om:value | only Service_Measure |
| | forService | only Service<br>min 1 Service |
| | gcii:consumedBy | only (serv:ServiceConsumer or serv:ServiceConsumerGroup) |

**Consumers**

The following classes from the GCI Shelter Ontology (Wang & Fox, 2015) define households for residential service consumers.

| Class | Property | Value Restriction |
|---|---|---|
| gcis:Household_size | owl:subclassOf | gci:GCI_quantity |
| | om:value | only Household_size_measure |
| | om:unit_of_measure | value gci:population_cardinality_unit |
| | prov:wasDerivedFrom | some cyc:census |
| gcis:Household_size_measure | owl:subclassOf | gci:GCI_measure |
| | om:unit_of_measure | value gci:population_cardinality_unit |
| | om:numeric_value | exactly 1 xsd:string |
| gcis:Average_houshold_size | owl:subclassOf | gci:GCI_quantity |
| | gci:for_city | only gci:City |
| | om:value | only Household_size_measure |
| gcis:Average_household_size_ Measure | owl:subclassOf | gci:GCI_measure |
| | pr:wasDerivedFrom | some Household_size_measure |
| | om:unit_of_measure | value gci:population_cardinality _unit |
| | om:numeric_value | exactly 1 xsd:string |

Similarly, information is required to describe non-residential service consumers. Note that these terms were created in (and then duplicated from) the building occupany ontology that will be discussed later in this section.

| Class | Property | Value Restriction |
|---|---|---|
| gcibo:Organization_size | owl:subclassOf | gci:GCI_quantity |
| | om:value | only Organization_size_measure |
| | om:unit_of_measure | value gci:population_cardinality_unit |
| gcibo: Organization_size_measure | owl:subclassOf | gci:GCI_measure |
| | om:unit_of_measure | value gci:population_cardinality_unit |
| | om:numeric_value | exactly 1 xsd:string |
| gcibo: Organization_Division_size | owl:subclassOf | gci:GCI_quantity |
| | om:value | only Organization_Division _size_measure |
| gcibo: Organization_Division_size_ Measure | owl:subclassOf | gci:GCI_measure |
| | om:unit_of_measure | value gci:population_cardinality _unit |
| | om:numeric_value | exactly 1 xsd:string |

With the above we can now define various types of consumers and households to capture energy consumption for the indicators:

| | | |
|---|---|---|
| ResidentialElectricalConsumer Household | owl:subclassOf | gcis:Household |
| | connectedthrough | some ElectricalServiceAccount |
| | consist_of | only ResidentialElectricalConsumer |
| ElectricalServiceAccount | owl:subclassOf | ServiceAccount |
| | hasServiceType | only ElectricalService |
| ResidentialElectricalConsumer | owl:subclassOf | ElectricalConsumer |
| | connectedthrough | some ElectricalServiceAccount |
| | consumes | only ElectricalService |
| | authorizedby | only ElectricalServiceProvider |

**Service Interruptions**

We extend our Service ontology to include Service Interruption events as required for Energy indicators 7.6 and 7.7.

The ServiceInterruption class specifies both the service accounts and service consumers impacted by an interruption. ISO37120 excludes all interruptions caused by extreme weather events so “causedbyWeather” was added so that extreme weather events can be filtered out. The num_accounts property identifies the number of ServiceAccount's affected by the interruption. These accounts may be listed in the impactAccount property.

| Class | Property | Value Restriction |
|---|---|---|
| ServiceInterruption | owl:subclassOf | lode:Event |
| | causedByWeather | only xsd:Boolean |
| | ot: HasDurationDescription | exactly 1 ot:DurationDescription |
| | impactAccount | only ServiceAccount |
| | num_accounts | exactly 1 xsd:nonNegativeInteger |
| | impactProvider | only ServiceProvider |
| ServiceInterruptionMeasure | owl:subclassOf | gci:GCI_measure |
| | om:unit_of_measure | value gci:interruption |
| serviceInterruptionVar | rdf:type | gs:Variable |
| | gs:has_Name | value ”num_accounts” |
| ServiceDurationMeasure | owl:subclassOf | gci:GCI_measure |
| | om:unit_of_measure | value om:hour |
| serviceDurationVar | rdf:type | gs:Variable |
| | gs:has_Name | value "HasDurationDescription" |

Within the service ontology, electricity specific concepts have been added as subclasses of the more general service classes. Specific units such as om:kilowatt_hour hours were imported from the measurement ontology.

| Class | Property | Value Restriction |
|---|---|---|
| ElectricalService | owl:subclassOf | Service |
| ElectricalConsumerGroup | owl:subclassOf | ServiceConsumerGroup |
| | gcii:consumes | some ElectricalService |
| ElectricalConsumer | owl:subclassOf | ServiceConsumer |
| | gcii:consumes | some ElectricalService |
| ElectricalServiceAccount | owl:subclassOf | ServiceAccount |
| | hasServiceType | only ElectricalService |
| ElectricalService ConsumptionMeasure | owl:subclassOf | ServiceConsumption Measure |
| | om:unit_of_measure | value om:kilowatt_hour |
| ElectricalService ConsumptionQuantity | owl:subclassOf | ServiceConsumption Quantity |
| | om:unit_of_measure | value om:kilowatt_hour |
| | forService | only ElectricalService |
| electricalConsumptionVar | rdf:type | gs:Variable |
| | gs:has_Name | value ”hasElectricalConsumption” |
| ElectricalServiceProvider | owl:subclassOf | ServiceProvider |
| | distributes | some ElectricalService |
| ElectricalServiceInterruption | owl:SubclassOf | ServiceInterruption |
| | forService | only ElectricalService |

**Service Production**

The following classes enable the representation of the different resources used to generate the electricity that it is distributing.

| Class | Property | Value Restriction |
|---|---|---|
| ServiceProduction | forService | some Service |
| Electrical_Service_Measure | owl:subclassOf | ServiceProductionMeasure |
| | om:unit_of_measure | value om:kilowatt_hour |
| Service_Production_Quantity | owl:sublclassof | gci:GCI_quantity |
| | forService | only Service |
| ElectricalServiceProduction Quantity | owl:subclassOf | Service_Production_Quantity |
| | unit_of_measure | value om:kilowatt_hour |
| | om:value | only ElectricalServiceProduction Measure |
| | forService | only ElectricalService |
| electricalProductionVar | rdf:type | gs:Variable |
| | | value quantityOfProduction |

The sources for electrical generation are defined as either renewables or non-renewable sources. Most terms are subclasses of Semanco (prefix sc) however, there are a few terms that were not defined such as “Tide” and “Wave.” The vocabulary was extended to add the property “QuantityOfProduction” to define what a grid mixture would look like for a city.

| Class | Property | Value Restriction |
|---|---|---|
| ElectricalPowerGenerationSource | owl:sublclassOf | ServiceProduction |
| | quantityOfProduction | Exactly 1 ElectricalService ProductionQuantity |
| NonRenewableSource | owl:sublcassOf | sem:Not_Renewable_Energy_Source |
| | owl:subclassOf | ElectricalPowerGenerationSource |
| NonRenewableSource | owl:subclassOf | sem:Renewable_Energy_Source |
| | owl:subclassOf | ElectricalPowerGenerationSource |
| Oil | owl:subclassOf | sem:Oil |
| | owl:subclassOf | NonRenewableSource |
| Natural_Gas | owl:subclassOf | sem:Natural_Gas |
| | owl:subclassOf | NonRenewableSource |
| Coal | owl:subclassOf | sem:Coal |
| | owl:subclassOf | NonRenewableSource |
| Nuclear | owl:subclassOf | sem:Nuclear |
| | owl:subclassOf | NonRenewableSource |
| Biomass | owl:subclassOf | sem:Biomass |
| | owl:subclassOf | RenewableSource |
| Hydro_Energy | owl:subclassOf | sem:Hydro_Energy |
| | owl:subclassOf | RenewableSource |
| Geothermal_Energy | owl:subclassOf | sem:Geotherhmal_Energy |
| | owl:subclassOf | RenewableSource |
| Solar_Energy | owl:subclassOf | sem:Solar |
| | owl:subclassOf | RenewableSource |
| Tide | owl:subclassOf | RenewableSource |
| Wave | owl:subclassOf | RenewableSource |

## 5.2 GCI Building Occupancy Ontology

The building occupancy ontology is designed to describe building types and the people who are residing in them. The GCI Building Occupancy ontology draws upon the GCI Shelter ontology (Wang & Fox, 2015) and the GCI Innovation ontology (Forde & Fox, 2015).

Some of the competency questions include:

1. (F) Is "Toronto Building 01" a residential building?
2. (F) What percentage of the floor space is used for residential purposes?
3. (F) How many households are in the city?
4. (F) What is the average number of people living in each household?
5. (F) How many households does "Toronto Building 01" have?
6. (F) What is the total population of residents the city?
7. (F) Which buildings in Toronto are owned by the government?
8. (F) How many public buildings are in the city?
9. (F) What is the floor area of all the public buildings?

dbepdia (prefix db) defines a building as a free standing structure that comprises of one or more rooms and spaces, covered by a roof, and enclosed with external walls or dividing walls that extend from the foundation to the roof. A Building has one main BuildingAddress. However, within the building may be multiple units that host different households and organizations. These tenants may have different services and accounts.

The definition of a "ResidentialBuilding" is regionally specific. The OECD (Residential Building, 2002) defines buildings with more than half of their floor space allocated to residential usage as residential buildings. The ResidentialBuilding class has specified the amount of Residental_FloorArea to ensure that the definitions match across data sets.

ISO37120 also states that PublicBuildings are defined as buildings that are owned by the government. By this logic, social housing would fit under the classes of ResidentialBuilding and PublicBuilding.

| Class | Property | Value Restriction |
|---|---|---|
| Building | owl:subclassOf | db:Building |
| | ic:hasAddress | exactly 1 ic:Address |
| | hasUnitAddress | some TenantSpace |
| | hasFloorArea | exactly 1 FloorArea_Quantity |
| | hasResFloorArea | exactly 1 FloorArea_Quantity |
| | occupied_by | some (gcis:Household or org:Division or org:Organization) |
| | hasElectricalConsumption | gcise:ElectricalServiceConsumption Quantity |
| | hasTenantSpace | only TenantSpace |
| | hasTenancy | only Tenant |
| | org:has_Ownership | exactly 1 org:Ownership |
| | owned_by | min 1 foaf:agent |
| CommericalBuilding | owl:subclassOf | Building |
| IndustrialBuilding | owl:subclassOf | Building |
| PublicBuilding | owl:subclassOf | Building |
| | has_Ownership | only org:GovernmentOrganization |
| ResidentialBuilding | owl:subclassOf | Building |
| | gcibo:hasHouseholds | min 1 gcis:Household |

Furthermore, the building occupancy ontology introduces the concept of TenantSpace to help capture that some buildings can be used by multiple sectors (residential buildings may have commercial store fronts.)

| Class | Property | Value Restriction |
|---|---|---|
| Tenant | owl:subclassOf | foaf:Agent |
| | occupies | min 1 TenantSpace |
| | represents | only (org:Organization or gcis:Household or org:Division) |
| gcis:Household | gcis:hasSize | only gcis:HouseHold_Size |
| | gcis:hasMember | only sch:Person |
| org:Organization | gcis:hasSize | only Organization_size |
| | consistsOf | only org:Division |
| | has_Ownership | only Ownership |
| | hasLegalName | 1 xsd:string |
| org:Division | divisionOf | some org:Organization |
| | gcis:hasSize | only Organization_division_size |
| TenantSpace | hasUnitIndicator | some rdfs:Literal |
| | insideBuilding | exactly 1 Building |
| | occupied_by | min 1 Tenant |
| | connectedServiceAccounts | min 1 ServiceAccount |

The energy indicators based on the number of residential users bases their information on household and average household size. These definitions were imported from the GCI Shelter Ontology (prefix gcis).

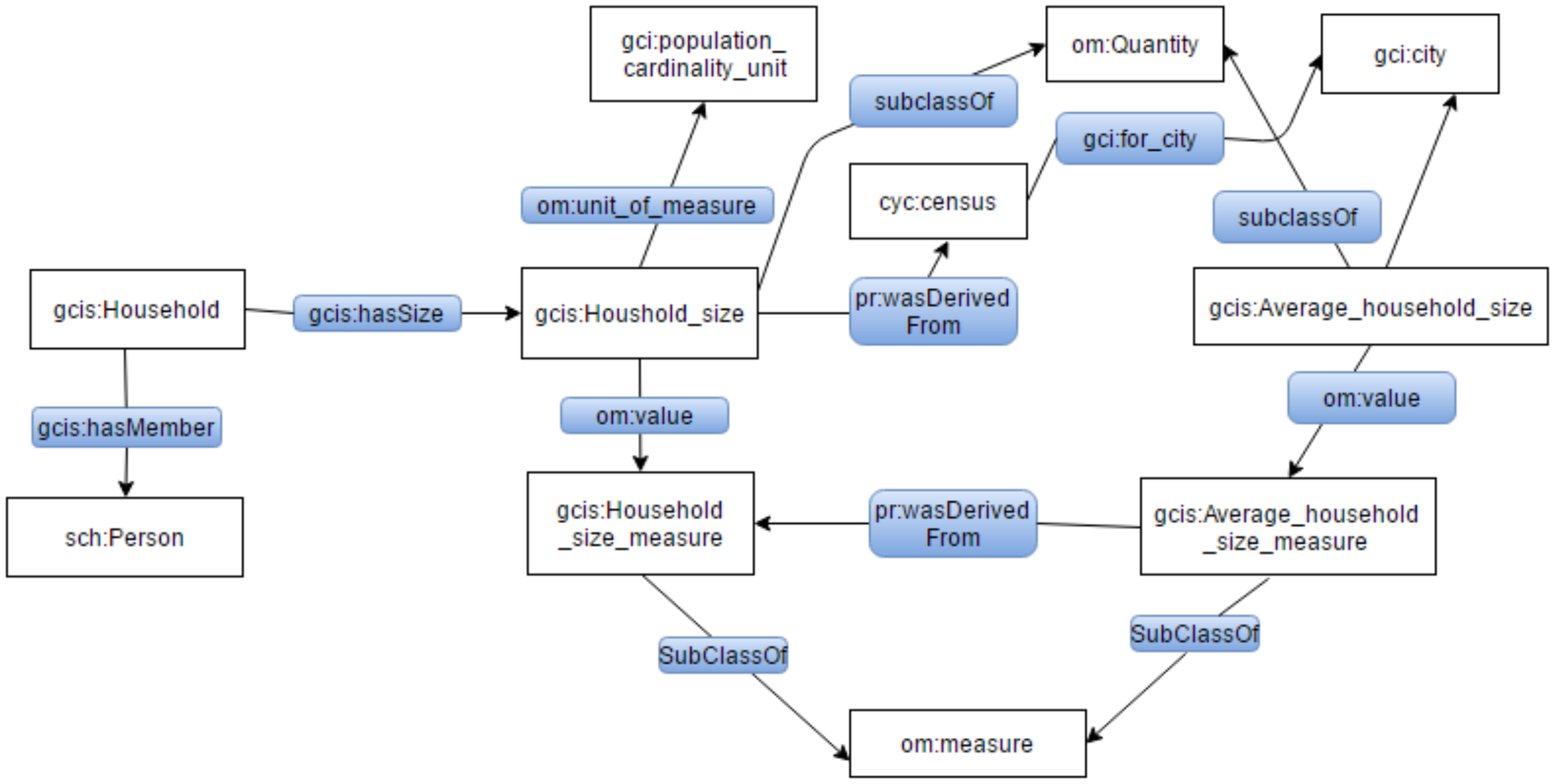

*Figure 5 Household Definition*

| **Class** | **Property** | **Value Restriction** |
|---|---|---|
| gcis:Household_size | owl:subclassOf | gci:GCI_quantity |
| | om:value | exactly 1 Household_size_measure |
| | om:unit_of_measure | value gci:population_cardinality_unit |
| | pr:wasDerivedFrom | some cyc:census |
| gcis:Household_size_measure | owl:subclassOf | gci:GCI_measure |
| | om:unit_of_measure | value gci:population_cardinality_unit |
| | om:numeric_value | exactly 1 xsd:string |
| gcis:Average_houshold_size | owl:subclassOf | gci:GCI_quantity |
| | gci:for_city | exactly 1 gci:City |
| | om:value | exactly 1 Household_size_measure |
| | unit_of_measure | value gci:population_cardinality_unit |
| gcis:Average_household_size_ Measure | owl:subclassOf | gci:GCI_measure |
| | pr:wasDervicedFrom | exactly 1 Household_size_measure |
| | om:unit_of_measure | value gci: population_cardinality _unit |
| | om:numeric_value | exactly 1 xsd:string |

The Organizational_size and Organizational_Division_size classes define the number of people who are occupying buildings for non-residential usage. Classes have been divided into Organization_size and Organization_division size because the full organization may or may not occupy the building.

| **Class** | **Property** | **Value Restriction** |
|---|---|---|
| Organization_size | owl:subclassOf | gci:GCI_quantity |
| | om:value | only Organization_size_measure |
| | om:unit_of_measure | value gci:population_cardinality_unit |
| Organization_size_measure | owl:subclassOf | gci:GCI_measure |
| | om:unit_of_measure | value gci:population_cardinality_unit |
| | om:numeric_value | exactly 1 xsd:string |
| Organization_Division_size | owl:subclassOf | gci:GCI_quantity |
| | om:value | exactly 1 Organization_Division _size_measure |
| Organization_Division_size_ Measure | owl:subclassOf | gci:GCI_measure |
| | om:unit_of_measure | value gci: population_cardinality _unit |
| | om:numeric_value | exactly 1 xsd:string |

The FloorArea_Quantity represents the quantity of the area of the floor within a building. This is associated to the total FloorArea_Measure with consist of the quantity of the floor area and the unit of measurement with is om:square_metre for this ontology.

| **Class** | **Property** | **Value Restriction** |
|---|---|---|
| FloorArea_Measure | owl:subclassOf | gci:GCI_measure |
| | om:unit_of_measure | value om:square_metre |
| FloorArea_Quantity | owl:subclassOf | gci:GCI_quantity |
| | om:unit_of_measure | value om:metre_Square |
| | om:value | only FloorArea_Measure |
| floorAreaVar | rdf:type | gs:Variable |
| | gs:has_Name | "hasFloorArea" |

# 6. GCI Foundation Ontology Infrastructure

In this section, we reviews the basic structure of a ratio indicator, its unit of measure, population and population size as defined in GCI foundational ontology (Fox, 2013), and upon which the Energy Indicators are based.

At the core of GCI Foundations Ontology is the OM measurement ontology (Rijgersberg et al. 2011). The purpose of a measurement ontology is to provide the underlying semantics of a number, such as what is being measured and the unit of measurement. The importance of grounding and indicator in a measurement ontology is to assure that the numbers are comparable, not that they are measuring the same thing, but they are measuring the same type ex. population of people with an authorized electrical connection and total city population are the same scale (ie. Thousands vs millions) and are in the same city.

Figure 6 depicts the basic classes of the OM ontology used to represent an indicator. There are three main classes in OM: a 'Quantity' that denotes what is being measured e.g. diameter of a ball; a 'Unit of Measure' that demotes how the quantity is measured, e.g., centimeters;

and a ‘Measure’ that denotes the value of the measurement which is linked to the both ‘Quantity’ and ‘Unit of Measure’. For example, a “percentage of population with authorization to electricity” is a subclass of “quantity” (gci:GCI_quantity) that has a value that is a subclass of “measure” whose units are a “percent” that is a type of “unit of measure”. The actual value measured is a property of the “measure” subclass “Population with authorization to electricity measures.”

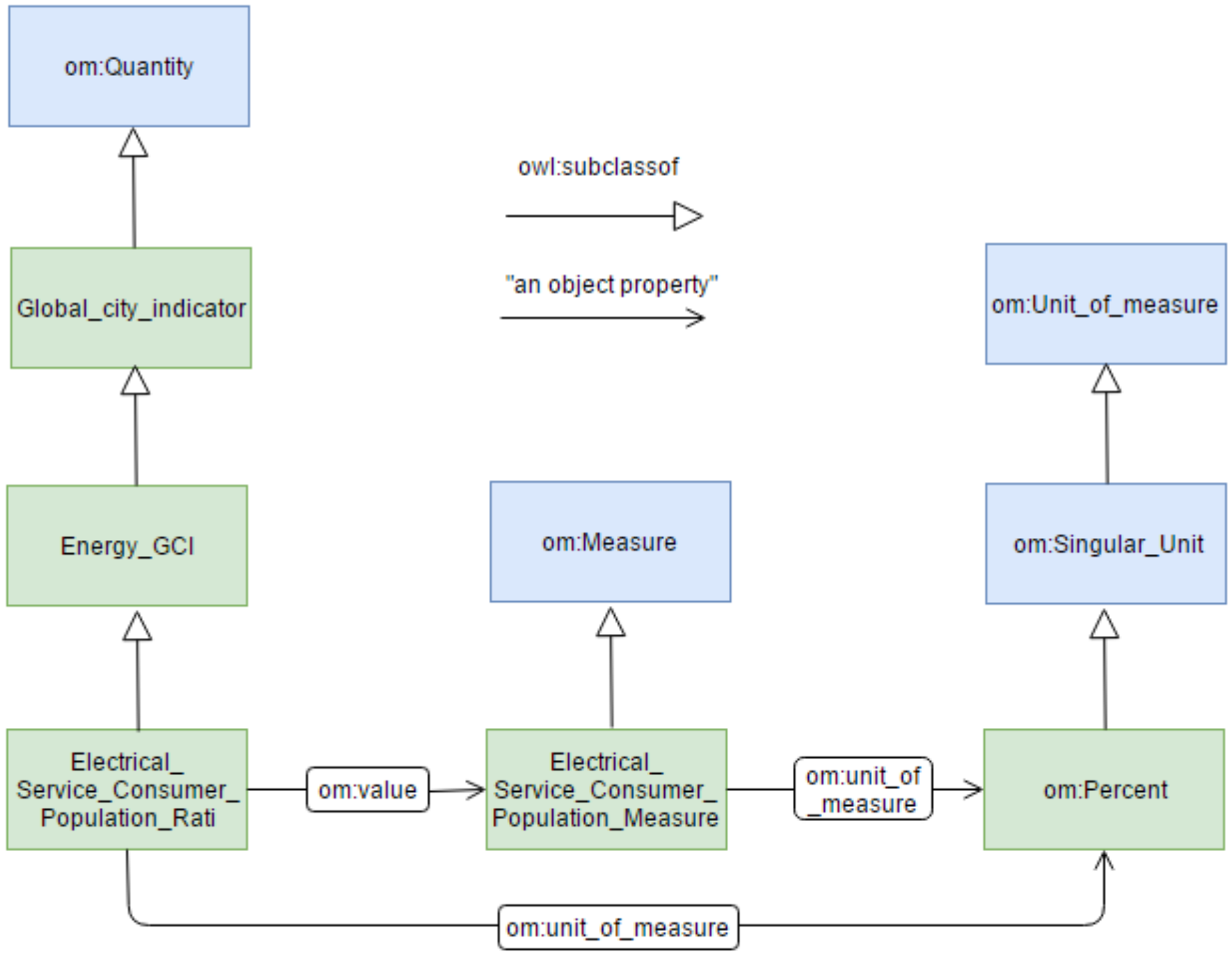

*Figure 6 Electrical Consumer Definition*

The number of people who are connected to electrical services is based on the population of connected households and the average size of a household in a city. One can view both as a statistical measurement in the sense that there is a population we want to perform a measurement of, a measurement being the count of the number of members that satisfy a description of an person connected to electrical service and a city resident, respectively. While the indicators require a count of members of the population, other measures may require statistics such as mean, standard deviation, etc.

All Energy indicators are ratio indicators (Fox, 2013). All ratio indicators have a numerator and denominator that are both represented by “population” class. A population is a collection of the same object such as people in a city and households. A ratio indicator has a unit of measure defined to be a “Population Ratio Unit” that specifies that the indicator is the ratio of the sizes (cardinalities) of two populations. One population size is the numerator and the other the denominator. A “Population Size” is defined as the cardinality of a “population,” and “Population” is defined by a “City” that the population is located in, and by a description of a “Person” within the “City” (Fox, 2013). For example, the “person” could be connected to electrical services (gcise:ElectricalServiceConsumer). Hence the population size (gci:Population_size) could be a number of “gcise:ElectricalServiceConsumer” in a particular city (gci:City). This general ontology structure is used in the indicator definitions outline in Section 7.

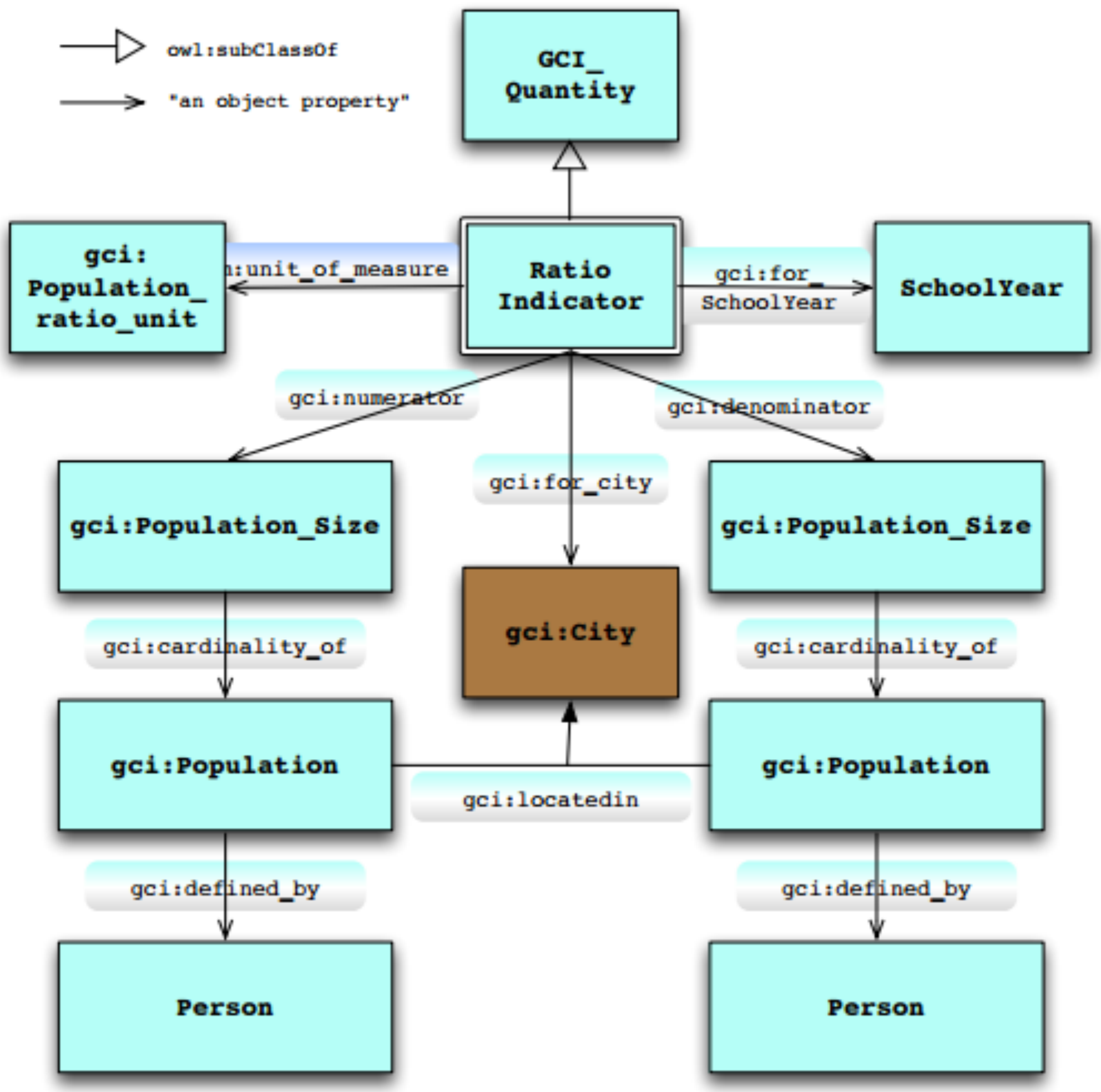

*Figure 7 Foundation Ontology Ratio Definition*

# 7. ISO 37120 Energy Indicators Definitions

The GCI Building Occupancy and GCI Service Ontologies provide the concepts needed to represent the definitions of ISO37120 Energy themed Indicators. This section provides a representation for each of the seven ISO37120 Energy Indicators' defintions. The OWL representation of the energy indicators can be found at http://ontology.eil.utoronto.ca/GCI/ISO37120/Energy.owl.

Common to multiple energy indicators such as 7.1, 7.2, and 7.5 is the city’s population. The class gci:City_Population_Size will be used as the denominator for the indicators 7.1, 7.2, and 7.5 and it is a gci:cardinality_of gci:City_Population which is a subclass of gci:Population.

| Class | Property | Value Restriction |
|---|---|---|
| gci:City_Population_Size | owl:subclassOf | gci:Population_size |
| | cardinality_of | exactly 1 gci:City_Population |
| | om:unit_of_measure | value gci:population_cardinality_unit |
| gci:City_Population | owl:subclassOf | gci:Population |
| | defined_by | exactly 1 gci:Resident |
| | function_of | some gcis:Average_household_size |
| | located_in | exactly 1 gci:city |

## 7.1 total residential electrical energy use per capita (kWh/year) (Core Indicator) (ISO 37120: 7.1)

This indicator is derived from the total amount of energy usage for residential buildings within city limits and divided by the city's population. The numerator is 7.1_Total_Residential_ElectricityConsumption_Quantity and the denominator is gci:City_Population_Size, as described above.

| Class | Property | Value Restriction |
|---|---|---|
| iso37120:7.1 | owl:subclassOf | iso37120:Energy |
| | om:denominator | exactly 1 gci:City_Population_Size |
| | om:numerator | exactly 1 7.1_Total_Residential_Electrical_Consumption_Quantity |
| | om:unit_of_measure | value gci:kwh_per_pc |

The 7.1_Total_Residential_ElectricityConsumption_Quantity is the defined as the total electrical usage in residential buildings within a city. To represent this, 7.1_Total_Residential_ElectricalConsumption_Quantity is a subclassOf gs:Sum which sums the total Residential_ElectricalConsumptionQuantities for the city's population of residential buildings. This population of residential buildings is defined by 7.1_Population_of_Residential_Buildings, which is defined by the population of residential buildings in a city.

| Class | Property | Value Restriction |
|---|---|---|
| 7.1_Total_Residential_Electrical_Consumption_Quantity | owl:subclassOf | gcise:ElectricalService ConsumptionQuantity |
| | om:value | exactly 1 gcise:Electrical_Service_Measure |
| | owl:subclassOf | gs: Sum |
| | gs:sum_of | exactly 1 7.1_Population_of_Residential_Buildings |
| | gs:sum_of_var | value gcise:electricalConsumptionVar |
| 7.1_Population_of_Residential_Buildings | owl:subclassOf | gs:Population |
| | gci:defined_by | only gcibo:ResidentialBuilding |

The following figure represents a tree format of the construction of iso37120:7.1.

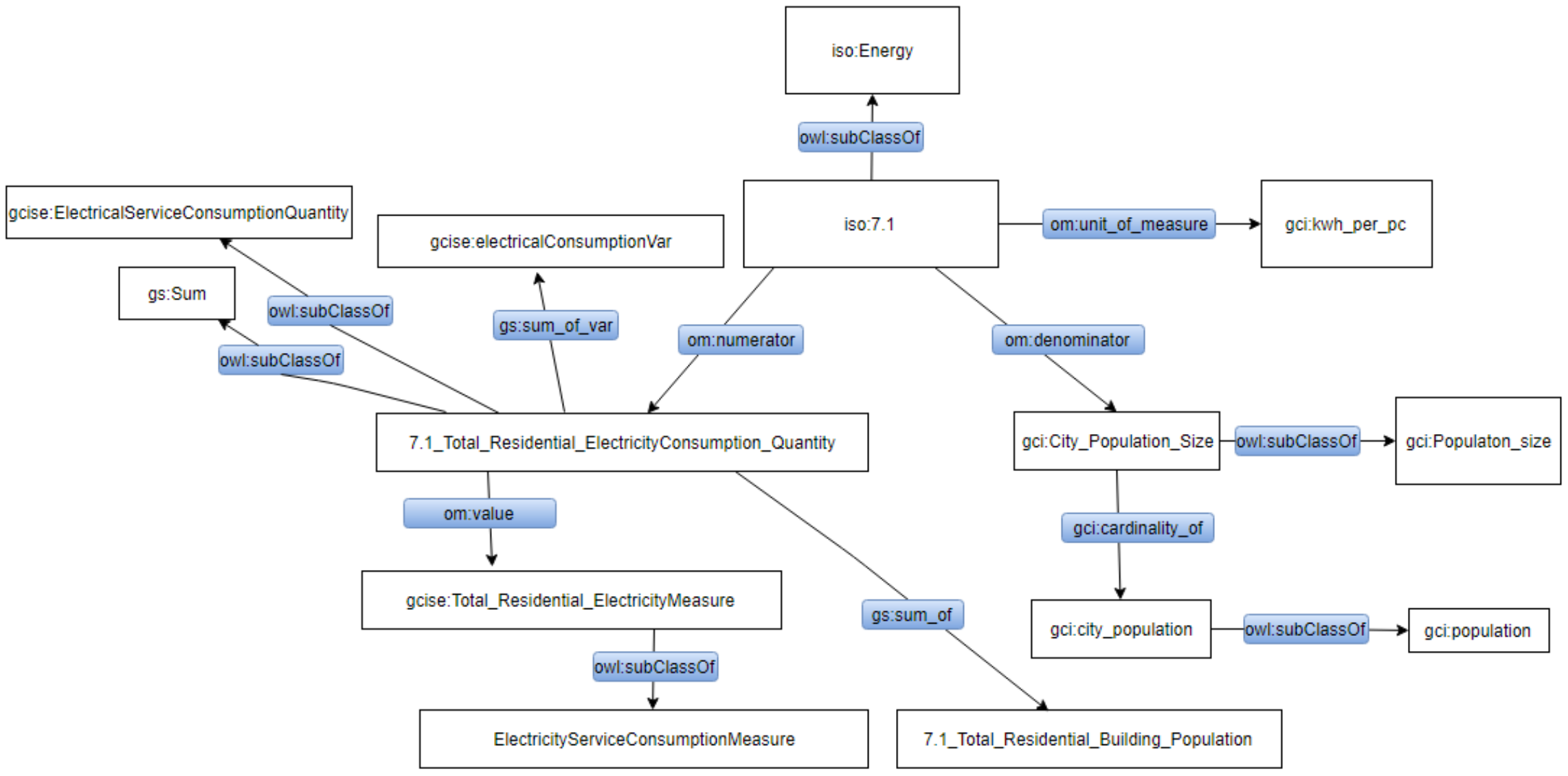

*Figure 8 ISO37120 7.1 Definition*

## 7.2 Percentage of city Population with authorized electrical service (core indicator) (ISO37120: 7.2)

This indicator describes the percentage of the total population who are legally accessing electrical services in their homes. Previously, we have defined the denominator of the indicator, which is the total population of the city. To construct the numerator of the indicator, we will define city residents who are receiving electrical services, if they are legally authorized to do so, and their total population. The numerator of the indicator is "7.2_Population_with_authorized_electrical_service."

| Class | Property | Value Restriction |
|---|---|---|
| iso37120:7.2 | owl:subclassOf | iso37120:Energy |
| | om:denominator | gci:City_Population_Size |
| | om:numerator | 7.2_Population_with_authorized electrical_service_size |
| | om:unit_of_measure | only om:percent |

The numerator shall be calculated as the number of households lawfully connected to the electrical grid multiplied by the average household size. The definitions for average household size have been imported from the GCI shelter ontology (prefix gcis). Although the calculation is done by multiplying the number of households with electrical service by the average number of average household population, the intent of the calculation to collect the total number of residential electrical consumers.

| Class | Property | Value Restriction |
|---|---|---|
| 7.2_Population_with_authorized | owl:subclassOf | gci:Product_Quantity |
| _electrical_service_size | om:unit_of_measure | value gci:population_cardinality_unit |
| | om:value | exactly 1 gci:Population_measure |
| | om:term_1 | exactly 1 gcis:Average_household_size |
| | om:term_2 | exactly 1 7.2_Population_of_Electrically ServicedHousehold_size |
| 7.2_Population_of_ElectricallyServicedHousehold_size | owl:subclassOf | gci:Population_size |
| | gci:cardinality_of | 7.2_Population_of_Electrically ServicedHouseholds |
| 7.2_Population_of_ElectricallyServicedHouseholds | owl:subclassOf | gci:Population |
| | gci:defined_by | only ResidentialElectrical ConsumerHousehold |

The definition of the numerator relies upon the definition of ResidentialElectricalConsumerHousehold in the energy service ontology (prefix gsise), defined earlier. Authorization arises from their having an account with an electrical service provider.

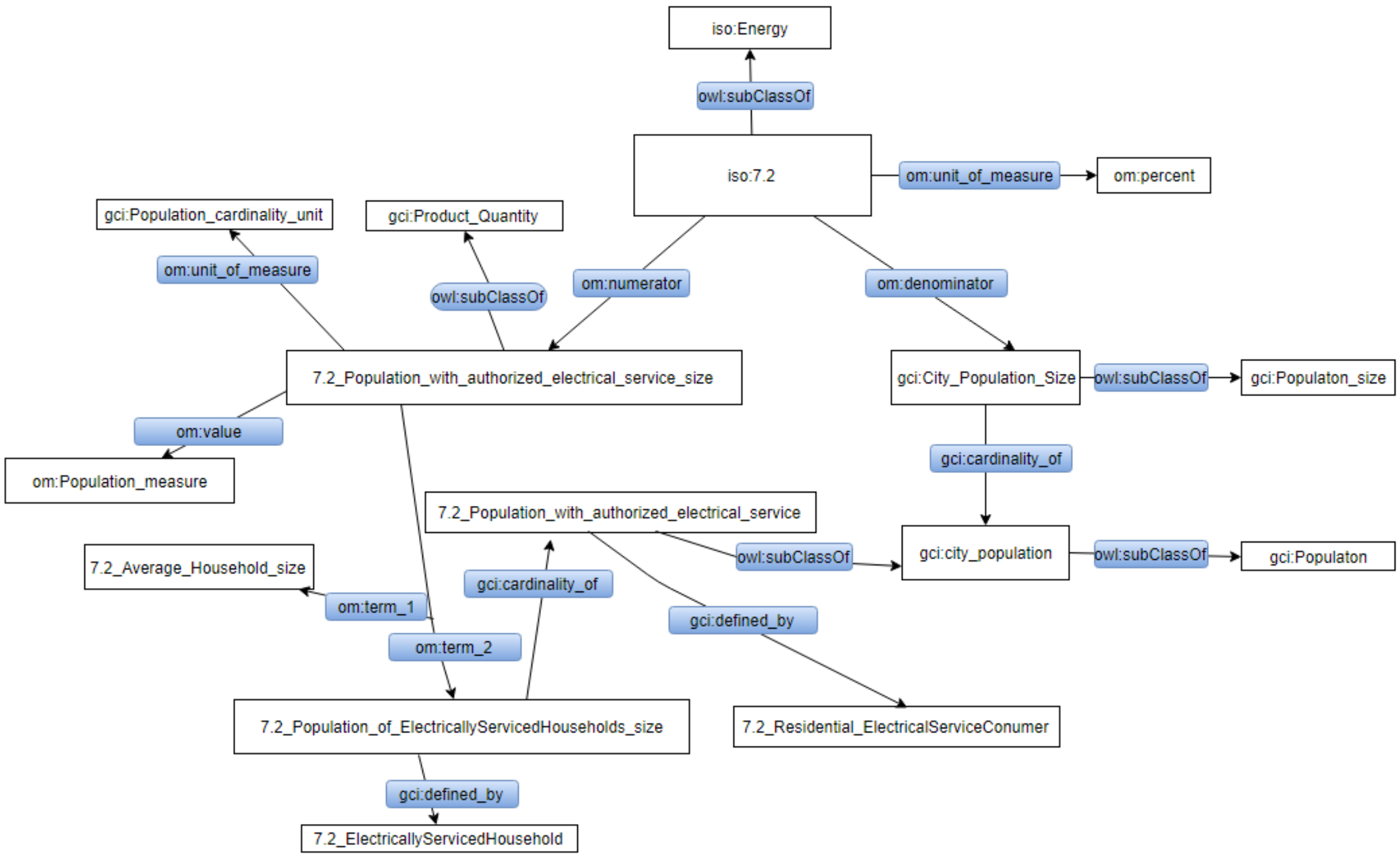

*Figure 9 ISO 37120 - 7.2 Definition*

## 7.3 Energy consumption of public buildings per year (kWh/m$^2$) (Core Indicator) (ISO 37120: 7.3)

The third indicator describes the total amount of electrical used in public buildings divided by the total amount of floor space in square metres. This indicator measures the energy efficiency of a city's public infrastructure. Public buildings have been defined in the Building Occupancy Ontology as buildings that are owned by government organizations.

First, we define the Iso37120:7:3 to have the numerator 7.3_Total_PublicBuilding_Electrical_Consumption_Quantity and it has the denominator of 7.3_Total_PulicBuilding_FloorSpace_Quantity.

| Class | Property | Value Restriction |
|---|---|---|
| iso37120:7.3 | owl:subclassOf | iso37120:Energy |
| | om:unit_of_measure | value gci:kwh_per_square_metre |
| | om:denominator | exactly 1 7.3_Total_PublicBuilding_ FloorSpace_Quantity |
| | om:numerator | exactly 1 7.3_ Total_PublicBuilding_Electrical Consumption_Quantity |

The total quantity of electrical consumption in public buildings is represented as the total amount of the sum of usage from the city's entire public building population.

| Class | Property | Value Restriction |
|---|---|---|
| 7.3_Total_PublicBuilding_Electrical_ Consumption_Quantity | owl:subclassOf | ElectricalService ConsumptionQuantity |
| | owl:subclassOf | gs:Sum |
| | gs:sum_of | exactly 1 7.3_Population_of_ Public_Buildings |
| | om:value | exactly 1 gcise:ElectricalService ConsumptionMeasure |
| | gs:sum_of_var | value gcise:electricalConsumptionVar |
| 7.3_Population_of_Public_ Buildings | owl:subclassOf | gs:Population |
| | gci:defined_by | only gcibo:PublicBuilding |

The total amount of floor space in a building is calculated similarly to the way that the total amount of electrical usage has been calculated. The amount of floor space is the sum of all of the floor space for the city's population of public buildings.

| Class | Property | Value Restriction |
|---|---|---|
| 7.3_Total_PublicBuilding_FloorSpac e_Quantity | owl:subclassOf | gcibo:FloorArea_Quantity |
| | owl:subclassOf | gs:Sum |
| | gs:sum_of | only 7.3_Population_of_Public _Buildings |
| | gs:sum_of_var | value gcibo:floorAreaVar |

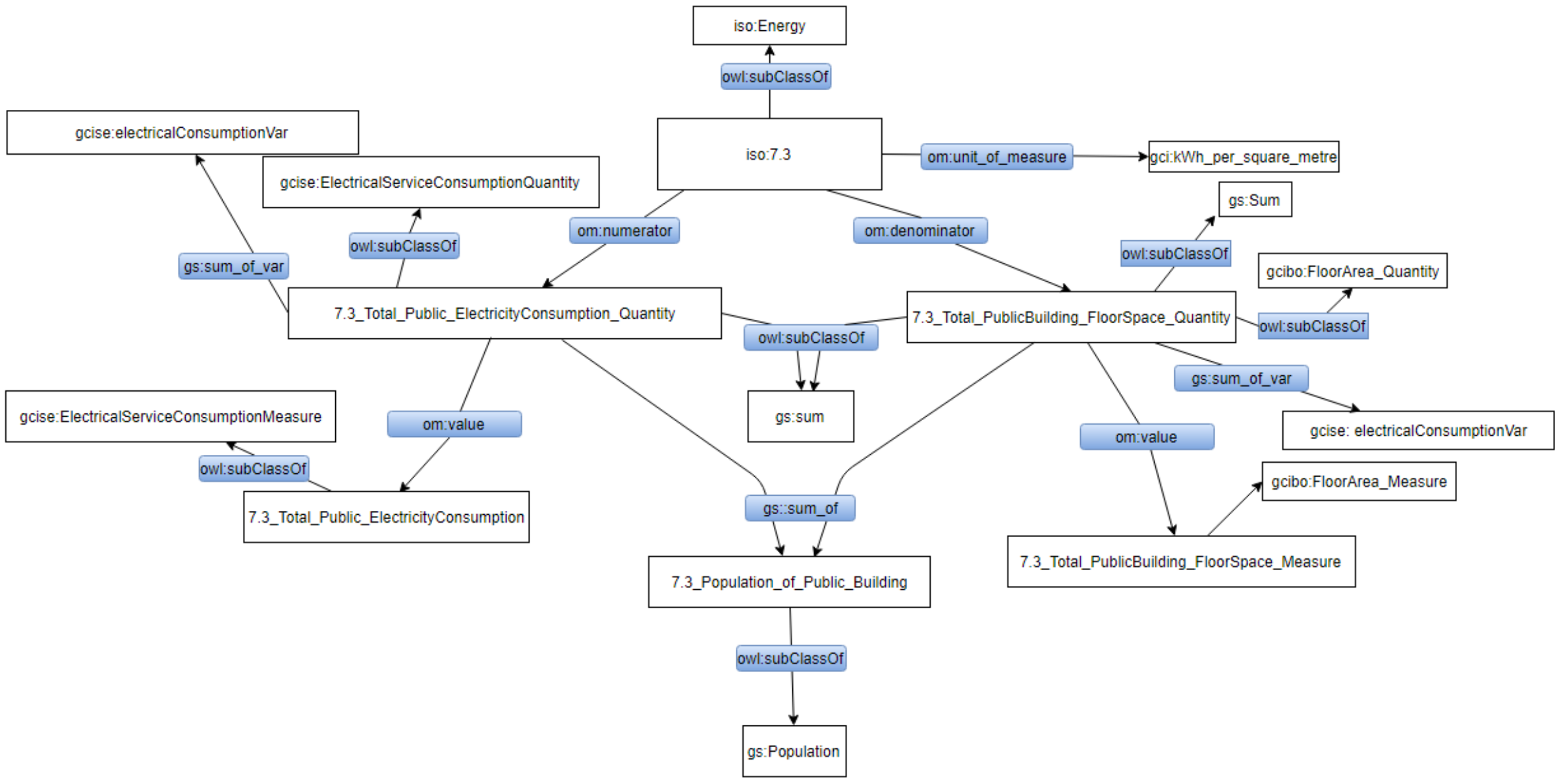

*Figure 10 ISO37120 7.3 Definition*

## 7.4 The percentage of total energy derived from renewable sources as a share of the city's total energy consumption (Core Indicator) (ISO37120:7:4)

The fourth core indicator evaluates the total amount of electricity that a city used that was initially produced by renewable energy sources. For this indicator, we estimate % of consumption by the ratio of renewable electrical production to total electrical production.

| Class | Property | Value Restriction |
|---|---|---|
| iso37120:7.4 | owl:subclassOf | iso37120:Energy |
| | om:denominator | exactly 1 7.4_Total_Electrical_ Production_Quantity |
| | om:numerator | exactly 1 7.4_Total_Electrical_ Production_From_Renewables _Quantity |
| | om:unit_of_measure | value om:percent |

First, the total amount of electricity produced by renewables shall be calculated by the amount of electricity generated by renewable sources including biomass, tide, solar, geothermal, wave, and wind that are nested under renewable sources. 7.4_Total_Electricity_Production_Quantity takes the sum of the gcise:QuantityOfProduction from each source in the population of renewable resources.

| Class | Property | Value Restriction |
|---|---|---|
| 7.4_Total_Electrical_Production_ From_Renewables_Quantity | owl:subclassOf | ElectricalServiceProduction Quantity |
| | owl:subclassOf | gs:Sum |
| | gs:sum_of | 7.4_RenewableSources |
| | gs:sum_of_var | value gcise:electricalProductionVar |
| | om:value | exactly 1 ElectricalServiceProductionMeasure |
| 7.4_RenewableSources | owl:subclassOf | gs:Population |
| | gci:defined_by | only (gcise:Biomass or gcise:Geothermal or gcise:Hydro_Energy or gcise:Solar_Energy or gcise:Tide or gcise:Wave or gcise:Wind_Energy) |

Next, we define the amount of electrical consumption by taking the sum of the electrical consumptions for the entire population of buildings in a city. This corresponds to the Variable “ElectricalServiceConsumption” for the data property gcise:Electricity_Consumption for each building.

| **Class** | **Property** | **Value Restriction** |
|---|---|---|
| 7.4_Total_Electrical_ Production_Quantity | owl:subclassOf | gcise:ElectricalServiceProductionQuantity |
| | owl:subclassOf | gs:Sum |
| | gs:sum_of | exactly 1 7.4_Production _Population |
| | om:value | exactly 1 ElectricalServiceProductionMeasure |
| | gs:sum_of_var | value gcise:electricalProductionVar |

The entire population of buildings in a city are summarized the loads taken from all buildings. For the ISO37120 indicators, public buildings and residential buildings are already defined.

| **Class** | **Property** | **Value Restriction** |
|---|---|---|
| 7.4_Total_Production_ Population | owl:subclassOf | gs:Population |
| | gci:defined_by | only gcise:ElectricalPowerGenerationSource |

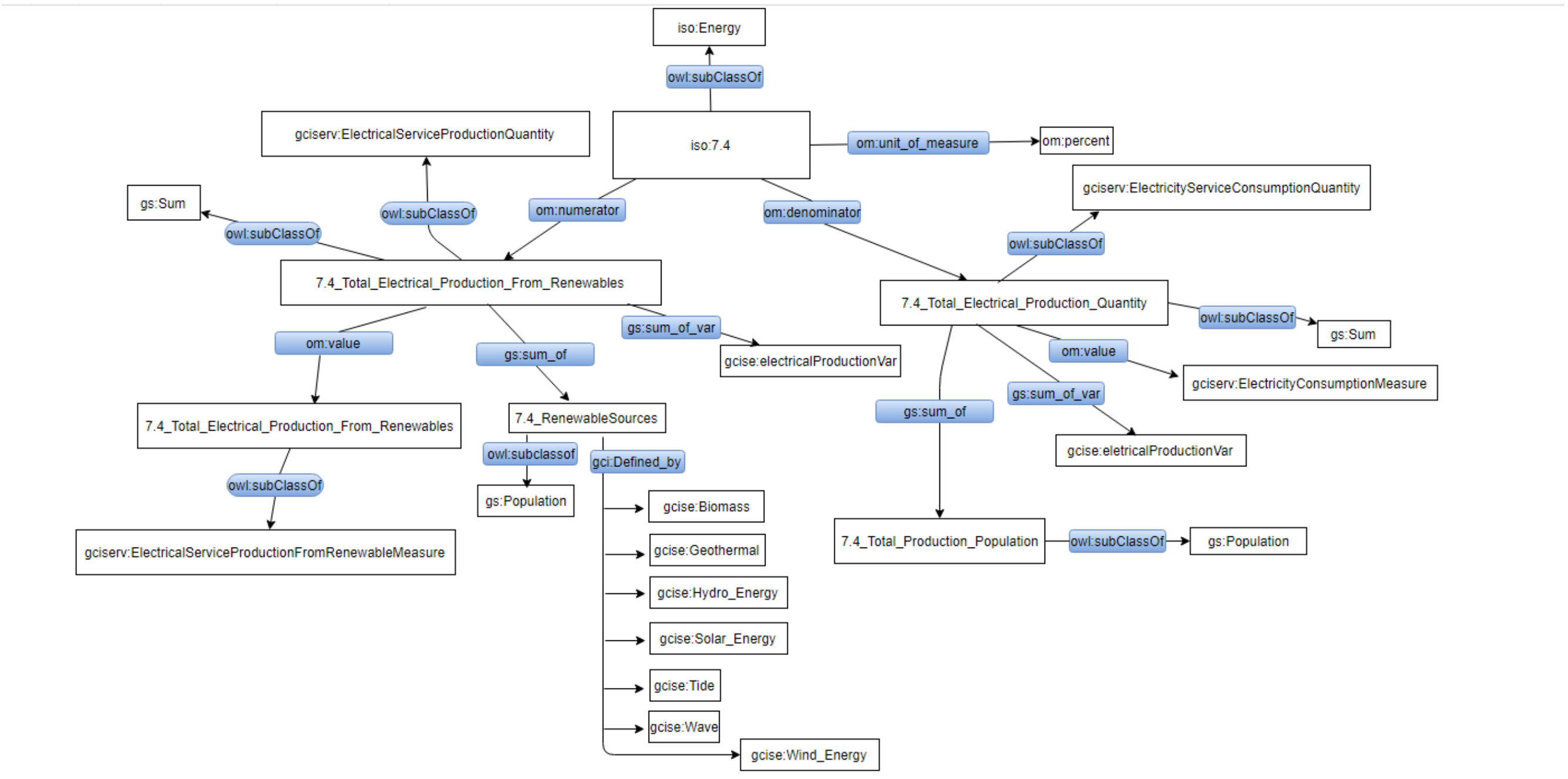

*Figure 11 ISO37120 7.4 Definition*

## 7.5 Total electrical energy use per capita (kWh/year) (Supporting Indicator) (ISO37120: 7.5)

The total amount of electrical usage per capita describes the total amount of electricity that a city would use (both residential and non-residential) per person. The numerator is similar to 7.4 and the denominator is the city population which was identified at the beginning of this section.

| Class | Property | Value Restriction |
|---|---|---|
| iso37120:7.5 | owl:subclassOf | iso37120:Energy |
| | om:unit_of_measure | value gci:kwh_per_pc |
| | om:denominator | exactly 1 gci:City_Population_Size |
| | om:numerator | Exactly 1 7.5_Total_Electrical_Consumption_Quantity |
| 7.5_Total_Electrical_ Consumption_Quantity | owl:subclassOf | ElectricalService ConsumptionQuantity |
| | owl:subclassOf | gs:Sum |
| | gs:sum_of | exactly 1 7.5_Total_Building_Population |
| | om:value | exactly 1 gcise:ElectricalService ConsumptionMeasure |
| | gs:sum_of_var | value gcise:electricalConsumptionVar |
| 7.5_Total_Building_Population | owl:subclassOf | gs:Population |
| | gci:defined_by | only gcibo:Building |

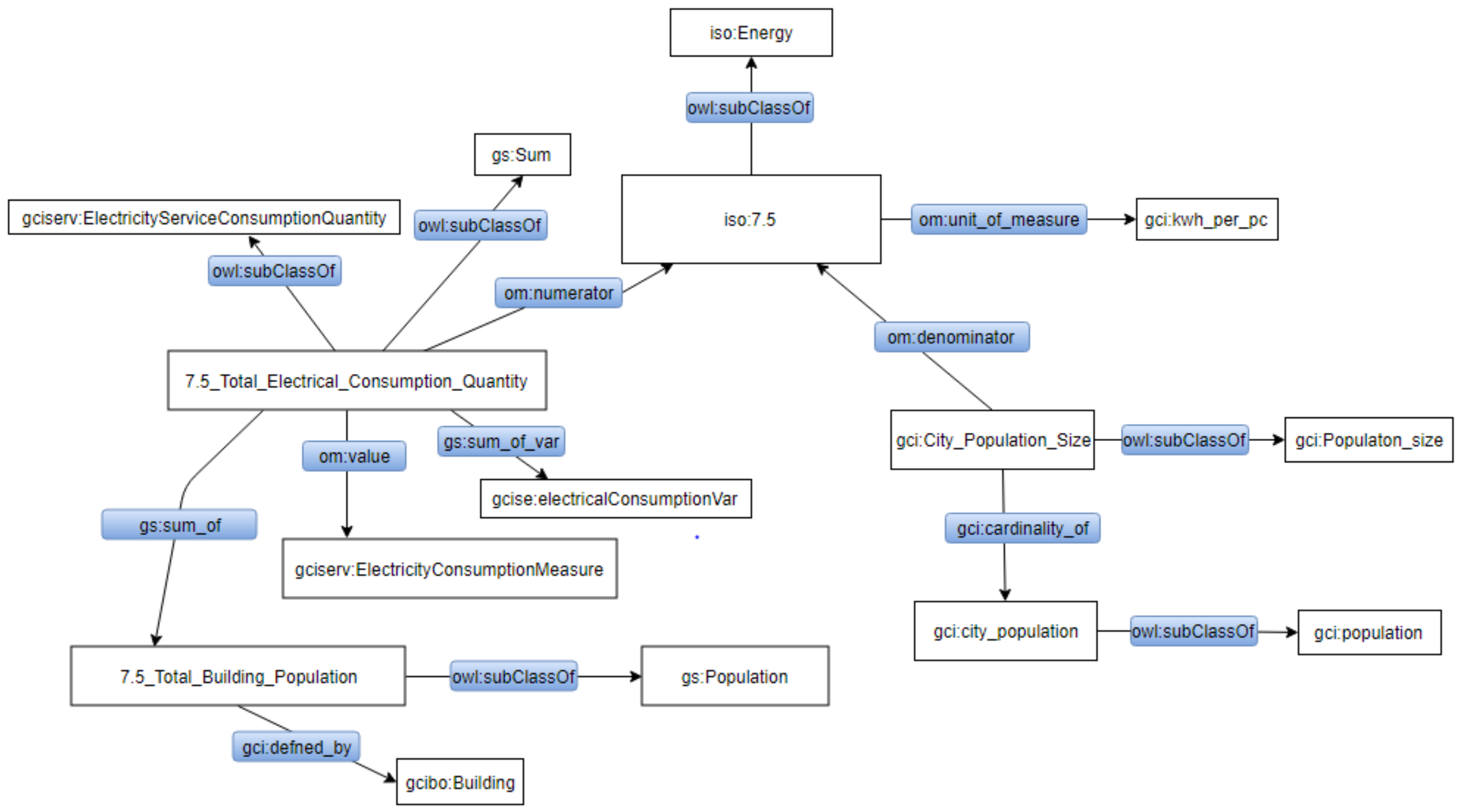

*Figure 12 ISO37120 Definition for 7.5*

## 7.6 Average number of electrical interruptions per year (Supporting Indicator) (ISO37120: 7.6)

The sixth indicator measures the consistency of a city's electrical service. The indicator seeks to derive the total number of electrical interruptions that each customer. Customer interruptions exclude service interruptions caused by extreme weather events.

| Class | Property | Value Restriction |
|---|---|---|
| iso37120:7.6 | owl:subclassOf | iso37120:Energy |
| | om:denominator | exactly 1 7.6_Customer_Account_Size |
| | om:numerator | exactly 1 7.6_Total_Count_of_Electrical_ Interruptions |
| | om:unit_of_measure | value gci:interruption_per_year |

The numerator sums the number of accounts that have been affected over the population of all electrical interruptions. The population excludes interruptions caused by weather.

| Class | Property | Value Restriction |
|---|---|---|
| 7.6_Total_Count_of_Electrical Interruptions | owl:subclassOf | gci:GCI_quantity |
| | owl:subclassOf | gs:Sum |
| | om:unit_of_measur e | gci:interruption |
| | om:value | exactly 1 gcise:ServiceInterruptionMeas ure |
| | gs:sum_of | Only 7.6_Population_of_Electrical Service_Interruptions |
| | gs:sum_of_var | value serviceInterruptionVar |
| 7.6_Population_of_ElectricalService_Interrupti ons | owl:subclassOf | gs:Population |
| | gci:defined_by | exactly 1 7.6_ElectricalServiceInterruptio n |
| 7.6_ElectricalServiceInterruption | owl:subclassOf | gcise:ElectricalService Interruption |
| | gcise:causedBy Weather | value xsd:false |

The denominator identifies the total number of customer accounts by defining a population composed of electrical service accounts.

| Class | Property | Value Restriction |
|---|---|---|
| 7.6_Customer_Account_Size | owl:subclassOf | gci:GCI_quantity |
| | owl:subclassOf | gs:Cardinality |
| | om:unit_of_measure | value gci:population_cardinality_unit |
| | om:value | exactly 1 gci:Population_measure |
| | gs:cardinality_of | exactly 1 |
| | | 7.6_Customer_Account_Pop |
| 7.6_Customer_Account_Pop | owl:subclassOf | gs:Population |
| | gci:defined_by | exactly 1 gcise:ElectricalServiceAccount |

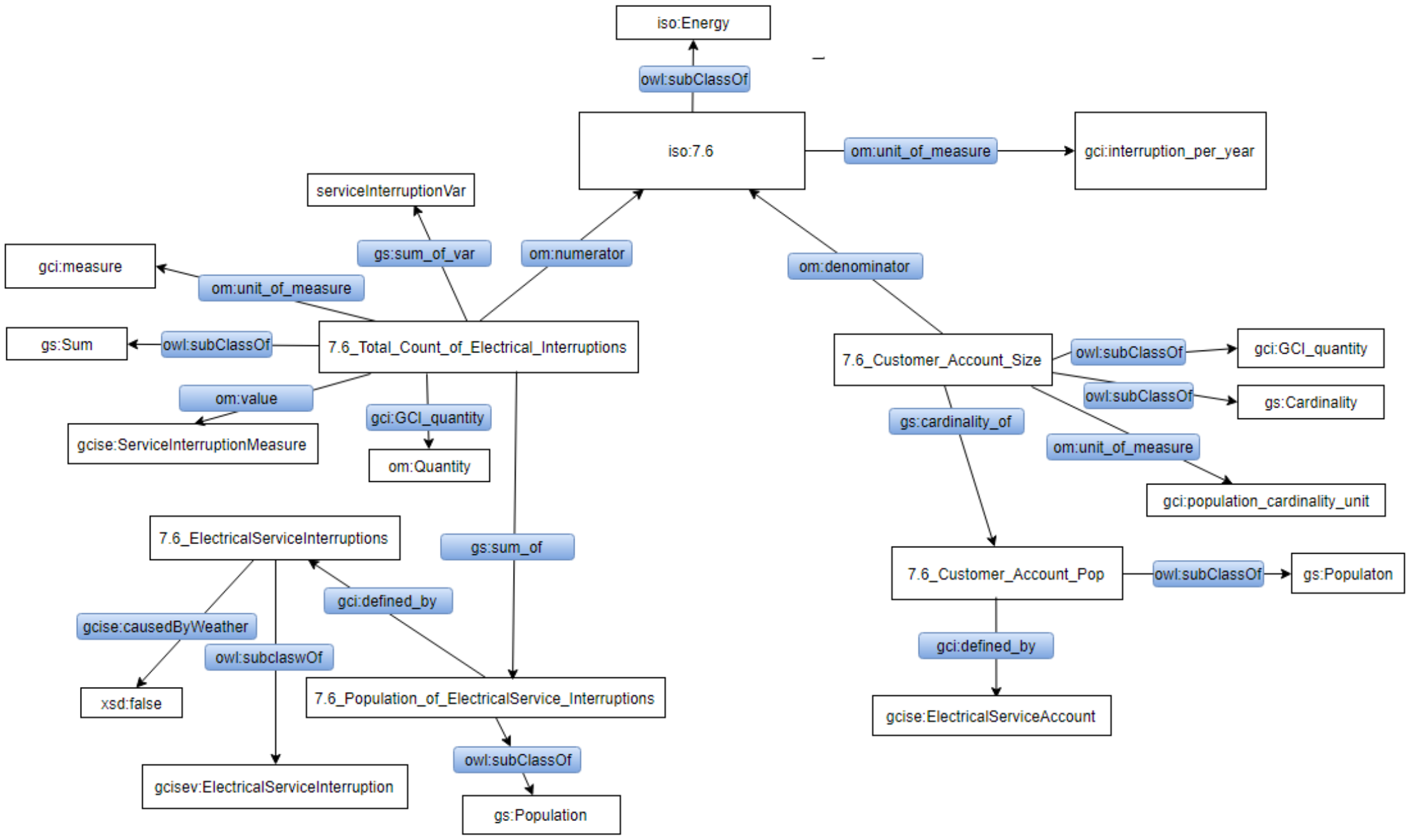

*Figure 13 ISO37120 - 7.6 Definition*

## 7.7 Average Length of Electrical Interruptions (Hours) (Supporting Indicator) (ISO37120:7.7)

The final energy theme indicator measures the average duration of the electrical interruptions. The indicator is derived dividing the sum of all the electrical interruptions divided by the total number of interruptions.

| Class | Property | Value Restriction |
|---|---|---|
| Iso37120:7.7 | owl:subclassOf | iso37120:Energy |
| | om:numerator | exactly 1 7.7_Sum_of_Duration_of_Electrical_Interruptions |
| | om:denominator | exactly 1 7.7_Total_Count_of_Electircal Interruptions |
| | om:unit_of_measure | om:hour |

We define the total duration of the electrical outages by taking the sum of the durations for all non-weather related interruptions.

| Class | Property | Value Restriction |
|---|---|---|
| 7.7_Sum_of_Duration_of _Electrical_Interruptions | owl:subclassOf | gci:Quantity |
| | owl:subclassOf | gs:Sum |
| | gs:sum_of | exactly 1 7.7_Electrical _Service_Interruption_Pop |
| | gs:sum_of_var | value gcise:serviceDurationVar |
| | om:value | exactly 1 gcise:ServiceDurationMeasure |
| | om:unit_of_measure | value om:hour |
| 7.7_Electrical_Service_Interruption_ Pop | owl:subclassOf | gs:Population |
| | gci:defined_by | exactly 1 7.7_Electrical_Service_Interruption |
| 7.7_Electrical_Service_Interruption | owl:subclassOf | gcise: ElectricalServiceInterruption |
| | gcise:causedByWeather | value xsd:false |

The denominator 7.7_Total_count_of_Electrical_Interruptions is defined as all electrical interruptions that were not caused by extreme weather events. It reuses the electrical service interruption population of the numerator but counts the size of the population

| **Class** | **Property** | **Value Restriction** |
|---|---|---|
| 7.7_Total_Count_of_Electrical Interruptions | owl:subclassOf | gci:GCI_quantity |
| | owl:subclassOf | gs:Cardinality |
| | om:unit_of_measure | value gci:interruption |
| | om:value | exactly 1 ServiceInterruptionMeasure |
| | gs:cardinality_of | exactly 1 7.7_Electrical_Service_Interruption_Pop |

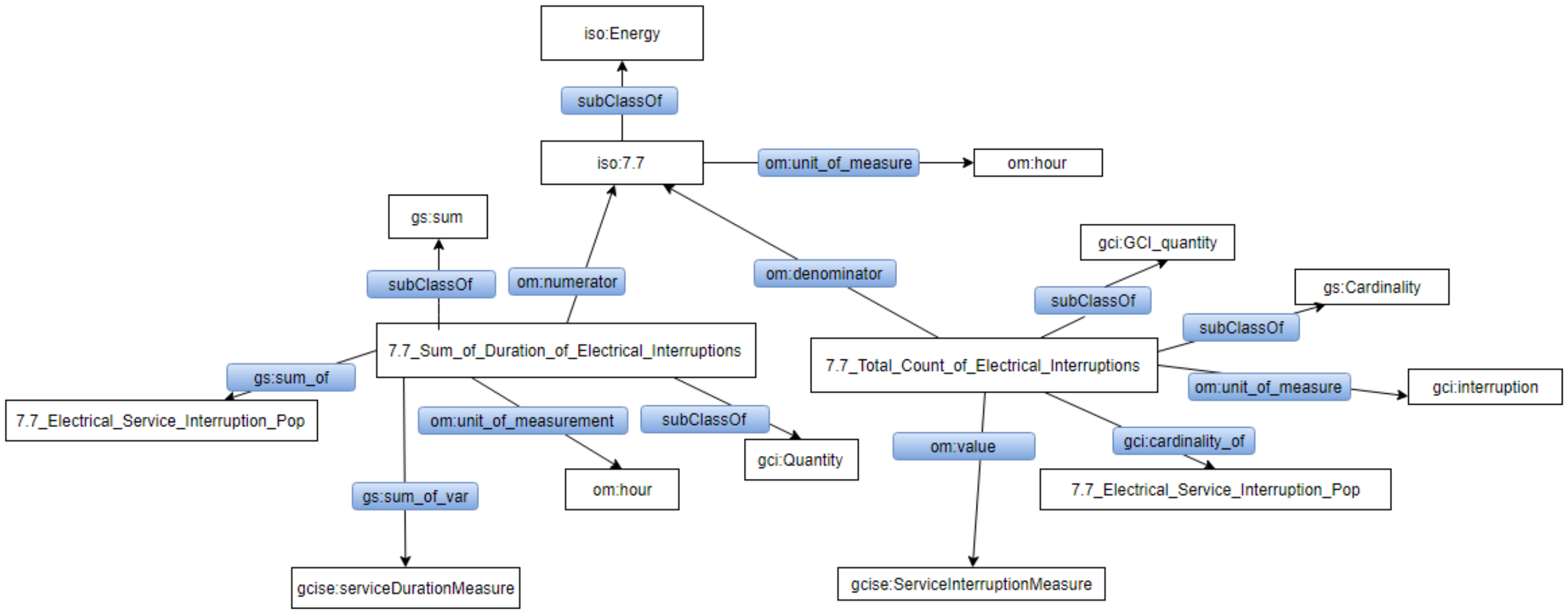

*Figure 14 ISO37120 7.7 Definition*

# 8. Evaluation

In this section we verify the Energy ontology by demonstrating that it can answer the competency questions. We use the City of Toronto in the Province of Ontario, Canada to answer the competency questions. In the following section, we define the example from the City of Toronto, using our ontology that will used to answer the competency questions. Appendix B contains the list of all prefixes.

| Instance | Property | Value |
|---|---|---|
| gn:6251999 | rdfs:label | Canada |
| | rdfs:type | gn:feature |
| | rdfs:type | sch:Country |
| gn:6093943 | rdfs:label | "Ontario" |
| | rdfs:type | gn:Feature |
| | rdfs:type | sch:Province |
| gn:6167865 | rdfs:Label | Toronto |
| | rdfs:type | gn:feature |
| | rdfs:type | sch:city |

| Instance | Property | Value |
|---|---|---|
| Toronto_Building_01 | rdfs:type | iso37120en:7.1_ResidentialBuilding |
| | ic:Address | Address_01 |
| | gcibo:hasFloorArea | Total_FS_TB01 |
| | gcibo:hasResFloorArea | Res_FS_TB01 |
| | gcibo:hasTenantSpace | gcibo:TenantSpace |
| | org:hasOwnerShip | org:privately_owned |
| | gcibo:owned_by | foaf:JohnDoe |
| Address_01 | rdfs:type | ic:Address |
| | ic:HasCity | gn:6167865 |
| | ic:hasState | gn:6093943 |
| | ic:hasCountry | gn:6251999 |
| | ic:Has_Street_number | 123 |
| | ic:Has_street | Fake |
| | ic:has_street_type | Street |
| | gcibo:hasBuilding | Toronto_Building_01 |
| Address_02 | rdfs:type | ic:Address |
| | ic:HasCity | gn:6167865 |
| | ic:hasState | gn:6093943 |
| | ic:hasCountry | gn:6251999 |
| | ic:has_Street_number | 14 |
| | ic:has_street | Carlton |
| | ic:has_street_type | Street |
| JohnDoe | rdfs:label | John Doe |
| | rdfs:type | foaf:agent |
| Total_FS_TB01 *(total floor area of building)* | rdfs:type | gcibo: FloorArea_Measure |
| | om:value | Total_FS_TB01_value |
| | om:unit_of_measure | om:square_metre |
| Total_FS_TB01_value | rdfs:type | gci:GCI_measure |
| | om:numerical_value | 1000 |
| | om:unit | om:square_metre |
| Res_FS_TB01 (*floor area in building for residential use)* | rdfs:type | gcibo: Res_FloorArea_Measure |
| | om:value | Total_FS_TB01_value |
| | om:unit_of_measure | om:square_metre |
| Res_FS_TB01_value | rdfs:type | gci:GCI_measure |
| | om:numerical_value | 900 |
| | om:unit_of_measure | om:square_metre |
| Toronto_Hydro | rdfs:type | gcise: ElectricalServiceProvider |
| | so:hasLegalName | 'Toronto Hydro' |
| | so:hasAddress | Address_02 |
| | gcise:distributes | gcise:ElectricalService |
| | gcise:Authorizes | ServiceAccount_01 |
| ServiceAccount_01 | rdfs:type | gcise: ElectricityServiceAccount |
| | gcise:accountActive | True |
| | gcise:Authorized_by | Toronto_Hydro |
| | gcise:hasServiceType | gcise:ElectricalService |
| | hasServiceAddress | Address_01 |
| | gcise:owned_by | JohnDoe |

| **Instance** | **Property** | **Value** |
|---|---|---|
| 7.1_ex *(Instance of 7.1)* | rdfs:type | iso37120:7.1 |
| | gci:numerator | 7.1_ex_Res_elec_ Consumption |
| | gci:denominator | Toronto_city_pop_size |
| | gci:for_city | gn:6167865 |
| | om:unit_of_measure | om:kwh_per_pc |
| | om:value | 7.1_ex_value |
| 7.1_ex_value *(value of 7.1)* | rdfs:type | gci:GCI_measure |
| | om:numerical_value | 1830 |
| | om:unit_of_measure | om:kwh_per_pc |
| 7.1_ex_Res_elec_ Consumption_quant *(numerator for 7.1)* | rdfs:type | iso37120en: 7.1_ Total_Residential_ Electrical_Consumption_ Quantity |
| | om:value | 7.1_ex_Res_elec_ Consumption |
| | om:unit_of_measure | om:Kilowatt_hour |
| | gci:for_city | gn:616765 |
| | gs:sum_of | 7.1_TO_Res_Build_Pop_ Value |
| | gs:sum_of_var | gcise:electricalConsumptionVar |
| 7.1_ex_Res_elec_ Consumption *(value of the numerator of 7.1)* | rdfs:type | gci:GCI_measure |
| | om:numerical_value | 5,073,000,000 |
| | om:unit_of_measure | kilowatt_hour |
| Toronto_city_pop | rdfs:type | gci:City_Population |
| | gci:located_in | gn:6167865 |
| Toronto_city_pop_size *(denominator for 7.1)* | rdfs:type | gci:City_Population_Size |
| | gci:cardinality_of | Toronto_city_pop |
| | om:value | Toronto_city_pop_size_ Value |
| | om:unit_of_measure | gci:population_cardinality_ unit |
| Toronto_city_pop_size_ Value *(value of denominator for 7.1)* | rdfs:type | gci:GCI_measure |
| | om:numerical_value | 2615000 |
| | om:unit_of_measure | gci:population_cardinality_ unit |
| 7.1_TO_Res_Build_Pop_ Size | rdfs:type | iso37120en:7.1_Total_Residential Building_Population |
| | gci:cardinality_of | 7.1_TO_Res_Build_Value |
| | om:unit_of_measure | gci:population_cardinality_ unit |
| 7.1_TO_Res_Build_Pop_ Value | rdfs:type | iso37120en:7.1_Population_of _ResidentialBuilding |
| | gci:GCI_measure | xsd:integer |
| | om:unit_of_measure | gci:population_cardinality _unit |

The following illustrates how the competency questions for ISO37120:7.1 are implemented in SPARQL.

1. (F) What city is the indicator for?

```
SELECT ?cityname WHERE
{7.1_ex gci:for_city ?city.
?city rdfs:label ?cityname}
```

   Answer: "Toronto"

2. (F) What is the total population of the city?

```
SELECT ?city ?city_pop_value WHERE
{?cityPop rdf:type gci:City_Population.
?cityPop gci:located_in ?city.
?cityPopSize gci:cardinality_of ?cityPop.
?cityPopSize om:value ?cityPopSize_measure.
?cityPopSize_measure om:numerical_value ?city_pop_value}
```

| City | City Population Size |
|---|---|
| Toronto | "2615000" ^^ xsd:integer |

3. (F) Is "Toronto_Building_01" a residential building?

```
SELECT ?BuildingType WHERE
{ Toronto_Builidng_01  owl:subclassOf ?BuildingClass }
```

   Answer: ResidentialBuilding

4. (CD) Who is the owners of Toronto_Building_01? What sector owns these buildings?

```
SELECT ?Owner ?Sector  WHERE
{ Toronto_Building_01 gcibo:owned_by ?Owner.
```

Toronto_Building_01 org:HasOwnership ?Sector}

| Owner | Sector |
|---|---|
| JonDoe | Privately_owned |

5. (F) What percentage of the floor space is used for residential purposes in Toronto_Building_01?

   SELECT (?Res_FS_Value/?Tot_FS_Value) AS ?percentage WHERE
   { Toronto_Building_01 gcibo:hasFloorArea ?Tot_FS.
   ?Tot_FS Om:value ?Tot_FS_Value.
   Toronto_Building_01 gcibo:hasResFloorArea ?Res_Tot_FS.
   ?ResPTot_FS om:value ?Res_Tot_FS_Value }

   Answer: 0.90

6. (F) How much energy was used per year in residential buildings?

   SELECT ?numeric_value ?unit WHERE
   {7.1_ex_Res_elec_Consumption om:value ?value.
   ?value om:numerical_value ?numeric_value.
   ?value om:unit ?unit }

| Numeric_value | Unit |
|---|---|
| 5,073,000,000 | Kilowatt_hour |

7. (F) What organizations provide electrical service in Toronto?

   SELECT ?ServiceProvider ?ServiceAccount WHERE
   { ?ServiceAccount a gcise:ServiceAccount.
   ?ServiceAccount gcise:hasServiceProvider gcise:ServiceProvider.
   ?ServiceAccount hasServiceAddress ?Address.
   ?Address ic:hasCity ?city.
   ?City rdfs:Label ?label.
   FILTER regex (?label. "Toronto")}

| ServiceProvider | ServiceAccount |
|---|---|
| Toronto_Hydro | ServiceAccount_01 |

8. (CI) Which service provider does each Toronto building use?

   SELECT DISTINCT ?ServiceProvider ?Building WHERE
   {?ServiceProvider a gcise:ServiceProvider.
   ?ServiceProvider gcise:authorizes ?ServiceAccount.
   ?ServiceAccount gcise:hasServiceAddress ?Address.
   ?Building a db:Building.

?Building ic:hasAddress ?Address }

| ServiceProvider | Building |
|---|---|
| Toronto_Hydro | Toronto_Building_01 |

9. (F) How many service accounts are there in residential_building_01?

SELECT (Count(?ServiceAccount) AS ?C) WHERE
{?ServiceAccount a gcise:ServiceAccount.
?ServiceAcccount gcise:hasServiceAddress ?Address.
?Address gcibo:hasBuilding Toronto_Building_01}

Answer: 1

# 9. Conclusions

The goal of this research was to define an ontology to represent the ISO37120 theme indicator definitions and the data use to derive a city's specific indicator value. In order to construct this ontology, we had to define some generic ontologies for energy related knowledge.

In summary, this research made the following contributions:

1.) Defines an Energy Ontology that covers a broader range of energy concepts related to total city level electrical usage and interruptions.
2.) Uses the above concepts to support and expand the definitions for indicators in "ISO 37120:7 Energy"
3.) Defines each of the ISO 37120: 7 Energy indicators using the foundation and GCI Service and Building Occupancy Ontology, providing a formalized computationally precise definition; and
4.) Provides a standard representation for general energy knowledge related to indicators, city specific versions of energy knowledge and the data used to derive the indicators value.

# 10. Acknowledgements

This research was supported in part by the Natural Science and Engineering Research Council of Canada.

# Appendix A – Key Ontologies

The Global City Indicator Foundation ontology can be found in:
http://ontology.eil.utoronto.ca/GCI/Foundation/GCI-Founation-v2.owl.

The Global City Indicator Service Ontologies for Service can be found in:
http://ontology.eil.utoronto.ca/GCI/Energy/GCI-Service.owl.

The Global City Indicator Service Ontologies for Building Occupancy can be found in:
http://ontology.eil.utoronto.ca/GCI/Energy/GCI-BuildingOccupancy.owl.

URIs for all of the ISO37120 indicators can be found in:
http://ontology.eil.utoronto.ca/ISO37120.owl.

Definitions of the ISO37120 Finance indicators, using the GCI Foundation and Energy ontologies can be found in:
http://ontology.eil.utoronto.ca/GCI/ISO37120/Energy.owl.

# Appendix B – Prefixes of Ontologies Used

| Prefix | Ontology | URL |
|---|---|---|
| db | dbpedia | http://dbpedia.org/ontology/ |
| foaf | FOAF | http://xmlns.com/foaf |
| gci | GCI foundation | http://ontology.eil.utoronto.ca/GCI/Foundation/GCI-Foundation.owl |
| gcibo | GCI Building Occupancy | http://ontology.eil.utoronto.ca/GCI/BuildingOccupancy/GCI-BuildingOccupancy.owl |
| gcii | GCI Innovation | http://ontology.eil.utoronto.ca/GCI/Innovation/GCI-Innovation.owl |
| gcis | GCI Shelter | http://ontology.eil.utoronto.ca/GCI/Shelters/GCI-Shelters.owl |
| gcise | GCI Service | http://ontology.eil.utoronto.ca/GCI/Energy/GCI-Service.owl |
| gn | Geonames | http://sws.geonames.org/ |
| gs | GovStat | http://ontology.eil.utoronto.ca/govstat.owl |
| ic | Icontact (international address Ontology) | http://ontology.eil.utoronto.ca/icontact.owl |
| iso37120 | iso 37120 IRIs | http://ontology.eil.utoronto.ca/ISO37120.owl |
| iso37120en | ISO 37120 Energy | http://ontology.eil.utoronto.ca/GCI/ISO37120/Energy.owl# |
| iso37120s | ISO 37120 Shelter | http://ontology.eil.utoronto.ca/GCI/ISO37120/Shelters.owl |
| lode | LODE Event | http://linkedevents.org/ontology/ |
| om | Measurement ontology | http://www.wurvoc.org/vocabularies/om-1.8 |
| org | TOVE organization | http://ontology.eil.utoronto.ca/organization.owl |
| ot | Owl time | http://www.w3.org/2006/time |
| pr | Prov | http://www.w3.org/ns/prov |
| sch | Schema.org | http://schema.org/ |
| sem | Semanco | http://semanco-tools.eu/ontology-releases/eu/semanco/ontology/SEMANCO/SEMANCO.owl |
| so | Service | http://purl.org/ontology/service |
| sumo | Suggested Upper Merged Ontology | http://www.ontologyportal.org/SUMO.owl# |